\documentclass[fleqn,usenatbib]{mnras}

\usepackage{newtxtext,newtxmath}

\usepackage[T1]{fontenc}
\usepackage{ae,aecompl}

\usepackage{graphicx}	
\usepackage{amsmath}	
\usepackage{amssymb}	
\usepackage{comment}	

\usepackage[usenames]{color}
\usepackage{xspace}

\newcommand{\mufasa}{{\sc Mufasa}}
\newcommand{\simba}{{\sc Simba}\xspace}

\newcommand{\HI}{\ion{H}{i}}

\newcommand{\hkpc}{h^{-1}{\rm kpc}}
\newcommand{\hmpc}{h^{-1}{\rm Mpc}}

\newcommand{\msolar}{\;{\rm M}_{\odot}}

\newcommand{\gizmo}{{\sc Gizmo}}
\newcommand{\caesar}{{\sc Caesar}}

\newcommand{\fHI}{f_{\rm HI}}
\newcommand{\fedd}{f_{\rm Edd}}

\newcommand{\rhalf}{R_{\rm half}}
\newcommand{\mstar}{\mbox{$M_\star$}\xspace}

\definecolor{Green}{rgb}{0.15,0.45,0.25}

\definecolor{mycolor}{rgb}{0,0,0}
\def\mycolor{\textcolor{mycolor}}

\title[Galaxy profiles and quenching]{The impact of quenching on galaxy profiles in the Simba simulation}

\author[S. Appleby et al.]
{\parbox[t]{\textwidth}{
Sarah Appleby$^{1}$\thanks{E-mail: sapple@roe.ac.uk},
Romeel Dav\'e$^{1,2,3}$, 
Katarina Kraljic$^{1}$, 
Daniel Angl\'es-Alc\'azar$^{4,5}$,\\
Desika Narayanan$^{6,7,8}$
}
\\
\\$^{1}$ SUPA\thanks{Scottish Universities Physics Alliance}, Institute for Astronomy, University of Edinburgh, Royal Observatory, Edinburgh EH9 3HJ, UK
\\$^2$ University of the Western Cape, Bellville, Cape Town 7535, South Africa
\\$^3$ South African Astronomical Observatories, Observatory, Cape Town 7925, South Africa
\\$^4$ Center for Computational Astrophysics, Flatiron Institute, 162 Fifth Avenue, New York, NY 10010, USA
\\$^5$ Department of Physics, University of Connecticut, 196 Auditorium Road, U-3046, Storrs, CT 06269-3046, USA
\\$^6$ Department of Astronomy, University of Florida, 211 Bryant Space Sciences Center, Gainesville, FL 32611, USA
\\$^7$ University of Florida Informatics Institute, 432 Newell Drive, CISE Bldg E251, Gainesville, FL 32611, USA
\\$^8$ Cosmic Dawn Center at the Niels Bohr Institute, University of Copenhagen\
 and DTU-Space, Technical University of Denmark
}

\date{Accepted XXX. Received YYY; in original form ZZZ}

\pubyear{2019}

\begin{document}
\label{firstpage}
\pagerange{\pageref{firstpage}--\pageref{lastpage}}
\maketitle

\begin{abstract}
We study specific star formation rate (sSFR) and gas profiles of star forming and green valley galaxies in the \simba\ cosmological hydrodynamic simulation. Star-forming galaxy half-light radii ($\rhalf$) at $z=0$ \mycolor{and their evolution ($\propto(1+z)^{-0.78}$) agree with observations. Passive galaxy $\rhalf$ agree with observations at high redshift, but by $z=0$ are too large, owing to numerical heating.} We compare \simba\ $z=0$ sSFR radial profiles for star forming and green valley galaxies to observations. \simba\ shows strong central depressions in star formation rate (SFR), sSFR, and gas fraction in green valley galaxies and massive star-forming systems, qualitatively as observed, owing to black hole X-ray feedback, which pushes central gas outwards. Turning off X-ray feedback leads to centrally peaked sSFR profiles as in other simulations. In conflict with observations, \simba\ yields green valley galaxies with strongly dropping sSFR profiles beyond $\ga \rhalf$, regardless of AGN feedback. The central depression owes to lowering molecular gas content; the drop in the outskirts owes to reduced star formation efficiency. \simba's satellites have higher central sSFR and lower outskirts sSFR than centrals, in qualitative agreement with observations.  At $z=2$ \simba\ does not show central depressions in massive star-forming galaxies, suggesting \simba's X-ray feedback should be more active at high-$z$. \mycolor{High resolution tests indicate central sSFR suppression is not sensitive to numerical resolution.}  Reproducing the central sSFR depression in $z=0$ green valley galaxies represents a unique success of \simba. The remaining discrepancies highlight the importance of SFR and gas profiles in constraining quenching mechanisms.
\end{abstract}



\section{Introduction}

Galaxies broadly fall into two classes: star-forming spiral galaxies, and quiescent elliptical galaxies. They occupy clearly distinct regions in the color-mass parameter space, the so-called `blue cloud' and `red sequence' \citep[e.g.][]{strateva_2001,baldry_2004,balogh_2004}. In between there is the `green valley' (GV), regarded as a transition zone since all galaxies must begin as star-forming while the most massive galaxies tend to be quiescent, which suggests that
at some point blue galaxies must stop forming stars and become red and dead \citep[e.g.][]{bell_2004,faber_2007,martin_2007,fang_2012}.
What physical driver(s) quench galaxies, i.e. transform them from being star-forming to quiescent, is a longstanding yet poorly understood question in galaxy evolution.

Modern galaxy formation models generally invoke feedback mechanisms associated with active galactic nuclei (AGN) to quench galaxies~\citep[see e.g.][and references therein]{somerville_2015}. Beyond this general notion, there remains much uncertainty regarding the physical mechanisms by which such AGN feedback operates, what triggers such feedback, and with which galaxy and/or halo properties such feedback most strongly correlates.  

Generally, quenching mechanisms fall into two broad categories. In merger quenching, major mergers are responsible for generating a starburst that evacuates the gas due to strong stellar and AGN feedback, leaving a dispersion-supported galaxy with little cold gas left to form stars~\citep{Springel_2005b,hopkins_2008}. In halo quenching, feedback associated with AGN causes the halo gas around the galaxy to be heated, which starves the central galaxy of further accretion, eventually causing a cessation of star formation~\citep{bower_2006,croton_2006,somerville_2008,gabor_2015,peng_2015}. Both models have observational support: for merger-driven quenching, observations clearly connect mergers with starbursts and AGN activity~\citep[e.g.][]{sanders_1996}, while for halo-driven quenching, bubbles seen in X-ray emission of galaxy clusters could potentially provide sufficient $PdV$ work to offset gas cooling~\citep{mcnamara_2007}. Many galaxy formation models, both semi-analytic and hydrodynamical, have implemented one or both of these forms of quenching in heuristic ways, and are thereby able to broadly reproduce the observed population of quenched galaxies.

A different set of constraints on quenching is provided by the bimodality in galaxy morphologies. At face value, merger quenching is attractive because it combines the quenching of star formation with a nearly concurrent transformation from spiral into elliptical. However, the existence of numerous ``red disks"~\citep{schawinski_2010,bundy_2010} with late-type morphologies but little or no star formation suggest that the morphological transformation and quenching are not necessarily coeval. Meanwhile, simulations suggest that after halo quenching causes starvation, the typically denser environment can result in minor mergers or galaxy harassment that can transform morphologies without the need for a major merger~\citep{oser_2012,gabor_2012}. However, the existence of rapidly-quenched systems such as post-starburst galaxies ~\citep[e.g.][]{zabludoff_1996,wild_2010} suggest that such a slow mechanism as starvation may not be sufficient to explain all quenched systems. Alternatively, it was shown that while mergers can lead to the formation of ellipticals, triggered AGN-regulated quenching is needed in order to freeze the post-merger morphology of a galaxy and prevent the disk re-formation \citep{gabor_2012,dubois_2016}. Hence it is likely that both quenching mechanisms are at play, with variations in importance that depend on galaxy mass, merger history, cosmic epoch, and environment.

To shed more light on galaxy quenching mechanisms, it is interesting to examine whether quenching occurs {\it inside-out} or {\it outside-in}, i.e. whether the bulge region drops in star formation rate prior to the disk, or vice versa.  Inside-out quenching could indicate some internal process is responsible for evacuating or heating the star-forming gas in the central region. \mycolor{Inside-out quenching can also be associated with `wet compaction' events due to minor mergers or tidal streams, leading to a ring of star-forming gas around the centre \citep{tacchella_2016b}.} Outside-in quenching might occur in particular if environmental processes such as gas stripping in the outskirts are the dominant quenching mechanisms.  A process such as starvation may slowly affect the entire disk, causing an overall drop in star formation everywhere \citep{van_den_Burgh_1991, elmegreen_2002}.  Thus by measuring the star formation rate and gas profiles of galaxies that are transitioning to being quenched, it may be possible to discriminate between quenching mechanisms.

Improving surveys can now measure the rate of galaxy growth via star formation as a function of galaxy radius, in massive galaxies that are likely to be on their way to being quenched.  Recently \citet{belfiore_2018} used spatially resolved spectroscopy from the Mapping Nearby Galaxies at APO (MaNGA) Sloan Digital Sky Survey (SDSS) \citep{bundy_2015} to derive star formation rates from H$\alpha$ flux and compute radial profiles of sSFR for star forming (SF) and GV galaxies. They find that at low stellar masses, the SF galaxies have flat radial sSFR profiles, but with increasing stellar mass galaxies show more centrally suppressed star formation. In particular, GV galaxies of all masses have sSFR profiles that are suppressed at all radii, as is expected from galaxies that are on their way to being quenched, and also show much stronger central suppression, particularly for galaxies with $\log (\mstar/\msolar) \ga 10.0$. In addition, decreasing SFR at the centre indicates that the suppression is not merely due to the increasing mass of the stellar bulge component, but is evidence for inside-out quenching. Similar findings in the literature show that transition galaxies with high stellar mass typically have central suppression in their sSFR profiles \citep{gonzalez_delgado_2016, coenda_2018, ellison_2018, liu_2018, sanchez_2018, spindler_2018, quai_2019}. Moreover, the fraction of inside-out quenching increases with stellar mass \citep{lin_2019}, suggesting that the fraction of inside-out quenching is higher than the fraction of outside-in quenching at a given stellar mass and environment.

At higher redshifts, SF galaxies can already be seen to develop central depressions in their SFR profiles as they begin their GV transition phase. At $z\approx 1$, SF galaxies with high mass $(10.5 < \log (\mstar/\msolar) < 11.0)$ show an enhancement in H$\alpha$, whereas less star forming galaxies of the same mass show central suppression of H$\alpha$ and inferred sSFR \citep{nelson_2016}. \mycolor{These observations are reproduced in the high-resolution Feedback In Realistic Environments (FIRE) zoom simulations, as a consequence of bursty star formation \citep{Orr_2017}.} At $z\approx 2$, dust corrected sSFR profiles are found to be flat for galaxies with $\log(\mstar/\msolar) < 11.0$, while for galaxies with $\log(\mstar/\msolar) > 11.0$, the sSFR profiles are centrally suppressed by a factor of $\sim$ 1 dex relative to the outskirts \citep{tacchella_2018}, demonstrating that inside-out quenching is already beginning in most massive star forming galaxies by $z\sim 2$. \mycolor{Inside-out quenching has also been independently observed via molecular gas profiles at $z\sim2$ \citep{spilker_2019}.} These observations support an inside-out quenching scenario, that is, the fractional rate of new star formation is higher in the outskirts than in the bulge region.

These observations of star formation distribution within galaxies provide strong constraints on galaxy formation models. Modern cosmological simulations are able to reproduce a variety of observational galaxy trends despite substantial differences in their prescriptions for sub-grid processes such as AGN feedback, motivating new tests by which to assess models. Recently \citet{starkenburg_2019} presented radial profiles of sSFR from the Evolution and Assembly of GaLaxies and their Environments \citep[EAGLE, ][]{crain_2015, schaye_2015} and Illustris \citep{genel_2014, vogelsberger_2014} cosmological simulations, both of which are able to quench galaxies in broad agreement with observations. They demonstrate that while the profiles of simulated SF galaxies are in reasonable agreement with observations \citep{belfiore_2018}, both simulations produce GV galaxies that have centrally concentrated star formation at all stellar masses, in direct contrast to observations. This suggests that galaxies in cosmological simulations are predominantly quenching from outside-in putatively owing to halo heating, and that current cosmological models have difficulty reproducing the observed inside-out quenching. This discrepancy between state-of-the-art cosmological simulations and observations identifies sSFR profiles as a key test for galaxy formation simulations.

In this paper, we examine the profiles of star formation rate and gas content, relative to stellar mass profiles, within the \simba\ simulation~\citep{dave_2019}.  \simba\ produces galaxies that are in good agreement with observations for a range of probes including stellar mass, star formation rate, neutral and molecular gas properties, black hole properties~\citep{thomas_2019}, and dust properties~\citep{li_2019}.  Most relevant for this work is that \simba\ yields a quenched fraction as a function of stellar mass that is in good agreement with observations~\citep{dave_2019}, hence it provides a useful platform to study how quenching proceeds within these simulated galaxies.  \simba\ includes three forms of AGN feedback, which heuristically describe radiative winds, bipolar jets, and X-ray radiation pressure, hence by running variants with these modules turned on and off, it becomes possible to examine which aspects of AGN feedback are responsible for quenching.  

This paper is organised as follows.  In \S\ref{sec:sims} we present the \simba simulations. \S\ref{sec:sizemass} presents the size-mass relation and its redshift evolution for simulated galaxies compared to observations.
In \S\ref{sec:radial} we show radial profiles for star-forming and GV galaxies, compare with the observed sSFR profiles, study the impact of different black hole feedback prescriptions on the radial profiles, study the differences in radial profiles between centrals and satellites, and examine the redshift evolution of radial profiles. Finally, in \S\ref{sec:summary} we conclude and summarize.

\section{Simulations}\label{sec:sims}

\simba, described more fully in \citet{dave_2019}, builds on its predecessor \mufasa\ \citep{dave_2016}. \simba\ is run using a modified version of the gravity plus hydrodynamics solver \textsc{Gizmo} \citep{hopkins_2015}, which uses the {\sc Gadget-3} tree-particle-mesh gravity solver \citep{springel_2005} and a meshless finite mass solver for hydrodynamics. In this work we use the fiducial 100 \(h^{-1}\) Mpc comoving volume, run from \(z=249\) down to \(z=0\) with 1024\(^3\) gas elements and 1024\(^3\) dark matter particles. For examining variations in the effects of different types of AGN feedback, we use 50\(h^{-1}\) Mpc comoving volumes with 512\(^3\) gas elements and 512\(^3\) dark matter particles. The mass resolution in both cases is \(9.6\times 10^7 M_{\odot}\) for dark matter and \(1.82\times 10^7 M_{\odot}\) for gas, and the minimum comoving gravitational softening length is \(\epsilon_{min} = 0.5h^{-1}\)kpc which corresponds to 0.5\% of the mean inter-particle spacing between the dark matter particles. \mycolor{To test numerical convergence, we use a higher resolution 25$\hmpc$ comoving volume with 512$^3$ gas elements and 512$^3$ dark matter particles. This simulation box has 8 times the mass resolution (\(2.3\times 10^6 M_{\odot}\) and \(1.2\times 10^7 M_{\odot}\) for the gas and dark matter particles, respectively) and twice the effective spatial resolution of the main \simba\ volume.} Cosmological initial conditions are generated using \textsc{Music} \citep{hahn_2011} and we assume a cosmology consistent with \citet{planck_collab_2016}: \(\Omega_M = 0.3\), \(\Omega_{\Lambda} = 0.7\), \(\Omega_{b} = 0.048\), \(H_0 = 68\) km s\(^{-1}\) Mpch\(^{-1}\) \(\sigma_8 = 0.82\), \(n_s = 0.97\). 

Star formation is modelled using an $H_2$-based \citet{schmidt_1959} relation, where the $H_2$ fraction is computed using the sub-grid prescription of \citet{krumholz_2011} based on metallicity and local column density, modified to account for variations in resolution \citep{dave_2016}. The star formation rate (SFR) is thus calculated from the density of molecular gas $\rho_{H_2}$ and the dynamical time \(t_{\textrm{dyn}}\) via \(\textrm{SFR}=\epsilon_* \rho_{H_2}/t_{\textrm{dyn}} \), where \(\epsilon_* = 0.02\) \citep{kennicutt_1998}.  The \HI\ fraction of gas particles is computed self-consistently within the code, accounting for self-shielding on the fly based on the prescription in \citet{rahmati_2013}, where the metagalactic ionizing flux strength is attenuated depending on the gas density, assuming a spatially uniform ionising background as specified by \citet{haardt_2012}.  This gives the total shielded gas, and subtracting off the molecular hydrogen fraction gives the fraction of gas in \HI.

Radiative cooling and photoionisation heating are implemented using the \textsc{Grackle}-3.1 library \citep{smith_2017}. The chemical enrichment model tracks 9 metals during the simulation, tracking enrichment from Type II supernovae (SNe), Type Ia SNe and asymptotic giant branch (AGB) stars, including locking some of the metals into dust. \simba\ includes star formation-driven galactic winds as decoupled, two-phase, metal-enriched winds with 30\% of the wind particles ejected hot and with a mass loading factor that scales with stellar mass, based on the FIRE \citep{Hopkins_2014} zoom simulation scalings from \citet{angles-alcazar_2017b}.

\begin{figure}
	\includegraphics[width=\columnwidth]{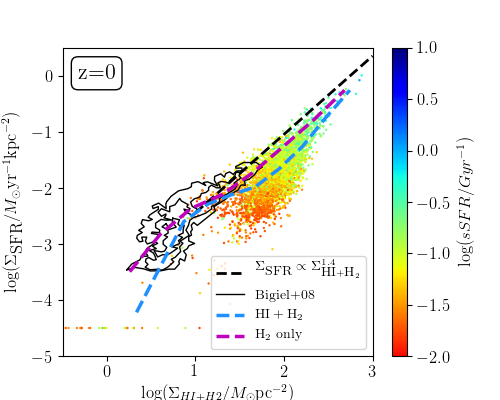}
    \caption{Surface density of gas ($\Sigma_{\HI+\textrm{H}_2}$) as a function of surface density of SFR ($\Sigma_{\textrm{SFR}}$) for SF and GV galaxies in the 100\(h^{-1}\) Mpc \simba box, colour coded by their sSFR. $\Sigma_{\HI+\textrm{H}_2}$ and $\Sigma_{\textrm{SFR}}$ are computed within the half-light radius of each galaxy (see  \S\ref{sec:sizemass} for details on how sizes are computed). Galaxies with $\Sigma_{\rm SFR}<10^{-4.5}\ \textrm{M}_{\odot}\textrm{yr}^{-1}\textrm{kpc}^{-2}$ have been plotted at that value for visibility.  A running mean for the non-quenched galaxies (sSFR$>10^{-1.8}$~Gyr$^{-1}$) is shown as the blue dashed line, and a running mean using only the molecular gas is shown as the magenta dashed line. The black dashed line is the best-fit relation to local spirals from \citet{kennicutt_1998}. The black contours are resolved galaxy observations from \citet{bigiel_2008}. The observations have been scaled to a Chabrier IMF as assumed in \simba.}
    \label{fig:kennicutt}
\end{figure}

Figure~\ref{fig:kennicutt} shows the Kennicutt-Schmidt (K-S) relation of SFR surface density versus \HI+H$_2$ surface density for \simba\ galaxies with $M_\star>10^{10}M_\odot$ at $z=0$. The points are colour-coded by sSFR.  We show a running \mycolor{mean} of this for all galaxies that have sSFR$>10^{-1.8}$~Gyr$^{-1}$ \mycolor{and non-zero gas surface density} as the blue dashed line. \mycolor{There are very few galaxies at low gas and SFR surface densities, so we display the mean values only if there are at least 5 galaxies in the given bin.} We show the observed \citet{kennicutt_1998} relation for star-forming galaxies (black dashed line), as well as the resolved relation from \citet{bigiel_2008} as the contours. \simba\ shows a reasonable agreement with the K-S relation, albeit slightly low in amplitude; this could be adjusted by increasing $\epsilon_*$.  Furthermore, it is seen that lower sSFR galaxies tend to lie below the K-S relation, which is consistent with observed early-type galaxies having lower star formation efficiencies~\citep[e.g.][]{DavisGreeneMa2016}.  The magenta dashed line shows a running median using only H$_2$, which shows a roughly linear relation between $\Sigma_{\rm SFR}$ and $\Sigma_{H2}$, and highlights how the turn-down in the K-S relation at low gas surface densities owes to an increase in the non-starforming \HI\ content.  Overall, \simba\ reproduces the K-S relation reasonably well, which shows that the relationship between gas and SFR surface density, central to the analysis in this paper, is adequately represented.

\mycolor{\simba tracks cosmic dust using a sub-resolution prescription, as a fraction of each gas element's metal budget that is passively advected with gas particles. The prescription is described in \citet{dave_2019} and broadly follows that in \citet{mckinnon_2016}.  Dust grains grow via condensation following \citet{dwek_1998} but with updated condensation efficiencies, as well as accretion of gas-phase metals via two-body collisions.  Dust is destroyed by collisions with thermally excited gas following the analytic approximation of dust growth rate from \citet{tsai_mathews_1995}. A mechanism for dust destruction via SN shocks (which enhance inertia and thermal sputtering of dust grains) is implemented following \citet{mckinnon_2016}. Dust is also instantaneously destroyed (dust mass and metals transformed into gas particles) in hot winds, during star formation, and in gas impacted by AGN feedback, except in cold star forming winds and radiative-mode Eddington AGN feedback to allow these winds to transport dust out of the galaxy. Dust that is destroyed is returned back to the gaseous metal phase.  \citet{li_2019} showed that \simba\ predicts global galaxy dust properties in reasonable agreement with observations across cosmic time.}

\simba's main improvement on \mufasa\ is the addition of black hole growth via torque-limited accretion 
\citep{hopkins_2011a, angles-alcazar_2013, angles-alcazar_2015} and AGN feedback via bipolar kinetic outflows. Black holes are seeded and grown during the simulation, and the accretion energy drives feedback that acts to quench galaxies. For cold gas (\(T <10^5\) K) black hole growth is implemented following the torque limited accretion model of \citet{angles-alcazar_2017a} which is based on \citet{hopkins_2011a}, while for hot gas (\(T >10^5\) K) Bondi accretion \citep{bondi_1952} is adopted. Unlike Bondi accretion, torque-limited accretion does not require the black hole to self-regulate its own growth~\citep{angles-alcazar_2015}, which allows for a more physical AGN feedback model.

The AGN feedback implementation in \simba\ is designed to mimic the observed dichotomy in black hole growth modes seen in real AGN \citep[e.g.][]{heckman_2014}: a `radiative' mode at high Eddington ratios (\(\fedd\)) characterised by mass-loaded radiatively-driven winds, and a `jet' mode at low \(\fedd\), characterised by high velocity jets of \(\sim 10^4\)~km~s\(^{-1}\). Our AGN outflow model has three modes of feedback: radiative, jet and X-ray. The radiative and jet modes are implemented kinetically, with outflows ejected in a direction \(\pm\) the angular momentum of the inner disk and with zero opening angle. We use variable velocity and mass outflow rate to mimic the transition between the radiative and jet modes when \(\fedd < 0.2\); full velocity jets are achieved when \(\fedd < 0.02\), and such outflows are heated to the halo virial temperature before ejection. For the radiative mode, particles are ejected without modifying their temperature at an outflow speed based on X-ray detected AGN in SDSS \citep{perna_2017}. For jet mode the outflow velocity increases as \(\fedd\) drops, capped at 7000 km s\(^{-1}\) above the radiative mode speed. We also require that \(M_{\textrm{BH}} > 10^{7.5}M_{\odot}\) to prevent small black holes with temporarily small accretion rates from driving high-powered jets. Finally, we include X-ray heating by black holes following the model in \citet{choi_2012}, which turns out to be quite important for our results. Our X-ray feedback implementation works in two ways: for non-ISM gas ($n_H<0.13$~cm$^{3}$), we directly increase the temperature of the gas, while for ISM gas half of the X-ray energy is used to give the gas particles a radial outwards kick, and the rest is added as heat.  As discussed in \citet{dave_2019}, globally the jet mode is primarily responsible for quenching galaxies, while the X-ray feedback has a small but important role in suppressing residual star formation, and radiative AGN feedback has little impact on galaxy properties.

Galaxies are identified using a 6-D friends-of-friends galaxy finder, using a spatial linking length of 0.0056 times the mean inter-particle spacing (equivalent to twice the minimum softening length), and a velocity linking length set to the local velocity dispersion. Black holes and \HI\ gas are assigned to the galaxy to which they are most gravitationally bound; we take the most massive black hole particle as the central black hole. Halos and galaxies are cross-matched using the \textsc{YT}-based package \caesar\footnote{caesar.readthedocs.io}, which outputs a catalog of pre-computed galaxy and halo properties. Particle data is read using \textsc{PyGadgetReader}\footnote{http://ascl.net/1411.001}. \simba\ outputs 151 snapshots from $z\approx 20\to 0$; here we employ snapshots at $z\approx 2$, $z\approx 1$ and $z=0$.

\section{Size-mass relation}\label{sec:sizemass}

Since we will scale our profiles by galaxy half-light radius, it is important to first check whether \simba\ yields sizes that are in reasonable agreement with observations.  For completeness, we do this at $z=0\to 2$, for star-forming and quenched systems, even though for the rest of this paper we will primarily focus on $z=0$ non-quenched galaxies.  

\begin{figure*}
	\includegraphics[width=\textwidth]{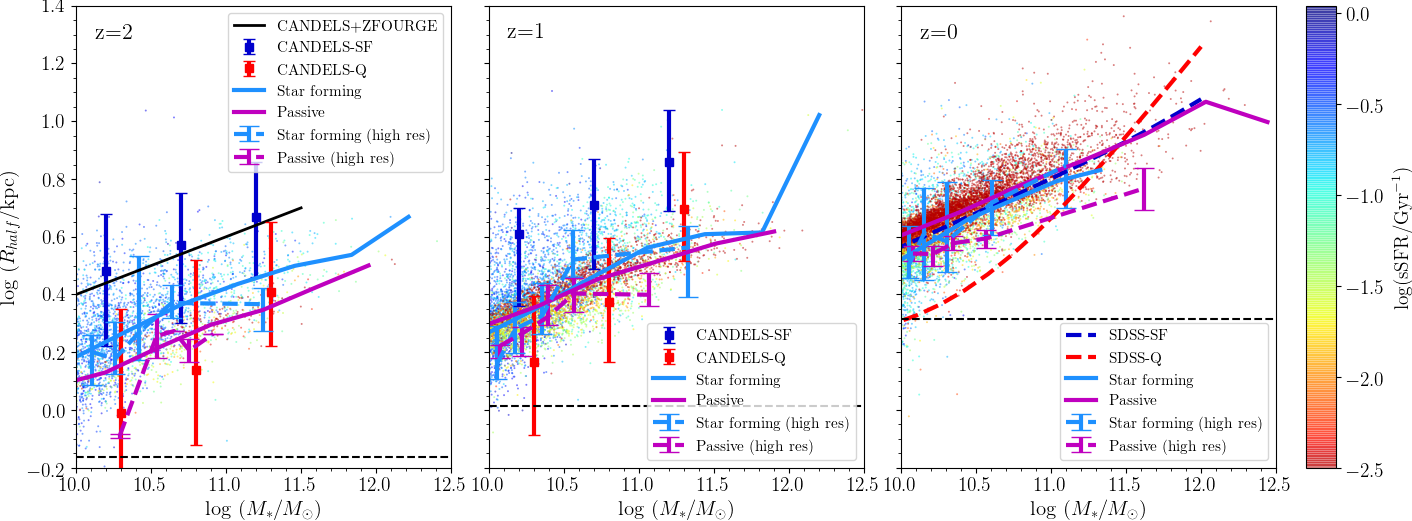}
	\vskip-0.1in
    \caption{Half-light radius as a function of stellar mass for $z$ = 2, 1, 0 (from left to right) for galaxies in the 100\(h^{-1}\) Mpc \simba box, colour coded by their sSFR. 
    The half-light radii are computed in the V band for $z$ = 2,1, and in the R band for $z$ = 0.
    The dark blue and magenta lines are the running medians for the star forming and passive galaxies respectively; \mycolor{the solid lines represent the main Simba volume, and the dashed lines represent the high resolution volume}. The horizontal dashed black lines show the effective size resolution limit of the fiducial simulation, below which galaxies are not well resolved. Observations are shown from \citet{van-der-wel_2014} and \citet{allen_2017} at z=2, \citet{van-der-wel_2014} at z=1 and \citet{zhang_2019} at z=0. 
    The sizes of the star forming galaxies are in broad agreement with the observations at $z=0$, while at higher redshifts they are smaller than observed. Passive galaxies have sizes in a good agreement with observations at $z=2$ and 1, but are larger than their star forming counterparts \mycolor{at $z=0$. Galaxies in the high resolution volume are consistent with the fiducial volume at all redshifts except for passive galaxies at $z=0$.}}
    \label{fig:halflight_all_z}
\end{figure*}

\mycolor{For each galaxy, we find $\rhalf$ by computing the half-luminosity radius in a particular band from individual stellar spectra of star particles using \textsc{Pyloser}\footnote{https://pyloser.readthedocs.io/en/latest/} (PYthon Line Of Sight Extinction by Ray-tracing). \textsc{Pyloser} generates a single stellar population (SSP) model using Flexible Stellar Population Synthesis\footnote{http://dfm.io/python-fsps/current/} \citep[{FSPS};][]{conroy_2009, conroy_2010} and uses this to compute a spectrum for each star particle, interpolated to its age and metallicity and assuming a \cite{chabrier_2003} IMF. Convolving the spectrum with a given bandpass gives the magnitude in a particular band. \textsc{Pyloser} accounts for dust attenuation by computing the extinction to each star particle based on the kernel-smoothed line of sight dust column density, converted to $A_V$ assuming Milky Way scalings \citep{watson_2011}. Given $A_V$, we attenuate each star's spectrum assuming a \citet{calzetti_2000} dust attenuation law for stars in galaxies with $\log $sSFR/Gyr$^{-1} > 0$, a Milky Way dust extinction law \citep{fitzpatrick_massa_2007} for $\log $sSFR/Gyr$^{-1} < -1$, and a linear combination of these in between; see \citet{salim_narayanan_2020} for a recent review of dust attenuation laws.  Given each star's (extincted) luminosity, we compute the half-light radius of every galaxy.} 

At $z=0$, we compute $\rhalf$ in the SDSS $r$ band to compare with SDSS data, and for higher redshifts we choose the V band to compare with that quoted from $k$-corrected CANDELS (Cosmic Assembly Near-infrared Deep Extragalactic Legacy Survey) data. The radius of each galaxy is found by averaging the three 2D projections along the \(x\), \(y\) and \(z\) axes, i.e. the sizes are computed along axes with random orientation with respect to the galaxies.

Figure \ref{fig:halflight_all_z} shows $\rhalf$ for all $M_\star>10^{10}M_\odot$ galaxies in the 100\(h^{-1}\) Mpc \simba\ volume at $z=2,1,0$ (left to right), colour coded by sSFR. We compare to \citet{van-der-wel_2014} and \citet{allen_2017} at $z=2$, \citet{van-der-wel_2014} at $z=1$, and \citet{zhang_2019} at $z=0$, separated into star forming and quiescent. At each redshift we show separate running medians for the SF and passive galaxy populations (magenta and blue lines), defining star forming as sSFR > \(10^{-1.8 + 0.3z}\)Gyr\(^{-1}\) as in \citet{dave_2019}. \mycolor{The effective spatial resolution is indicated by the dashed black lines in each panel. This is the radius out to which the gravitational force is softened, given by the minimum Plummer softening scale ($0.5 \hkpc$, comoving) multiplied by a factor of 2.8 for our assumed cubic spline kernel \citep{springel_2005}.}

At $z=0$, the sizes of the SF galaxies are in good agreement with the observations. The good agreement with data for the SF galaxies is an important success for \simba; there was no tuning done to obtain this agreement. In contrast, passive galaxies have sizes that are significantly larger than the observations at $\mstar\la 10^{11.5} M_\odot$.  In fact, the passive galaxies are \mycolor{slightly} larger than the SF galaxies at all masses, which is the opposite of what is seen for real galaxies. This indicates that we are not reproducing the compact nature of the stellar distribution in passive galaxies, particularly at low masses. This was already noted at $z=0$ in \citet{dave_2019}.  \mycolor{At $z=1$ and $z=2$, we see that passive galaxy sizes are in better agreement with observations, but here the star-forming galaxies are too small.} We have checked that we see the same trends in the simulations without X-ray and/or jet feedback, showing that this does not owe to the AGN feedback model in \simba.

By examining stellar surface density images of \simba\ galaxies, we have noticed that our SF galaxies do not have the extended thin stellar disks that are common to real SF galaxies. They typically have a gas component that has settled into a thinner disk, but a much puffier thick disk or even spheroidal stellar distribution. Unlike stars, gas particles in the simulation are able to dissipate energy through hydrodynamic interactions, allowing them to settle into disks more easily than the stellar component. Since the $r$ or V band half-light radii of the galaxies generally trace the stellar component, this indicates that something is puffing out stellar orbits. We have checked that newly-formed stars lie in a thin disk.

One possibility is numerical resolution, as older stars in present-day galaxies have undergone dozens of orbits where two-body effects and other dynamical noise can artificially heat the orbits. \mycolor{We investigate this by looking at the higher resolution $25 \hmpc$, $2\times512^3$ particle \simba volume. The dashed lines in Figure \ref{fig:halflight_all_z} show that at higher resolution, the sizes of star forming galaxies and the high redshift passive galaxies agree with the sizes of the lower resolution box, showing that these sizes are numerically converged.
This is not the case, however, for the passive galaxies at $z=0$, where the increased resolution has decreased the sizes, particularly at high stellar masses. This indicates that the large sizes of passive galaxies in the main simulation is likely due to numerical heating of stellar orbits, since these galaxies are composed almost entirely of star particles, with little gas. The fact that this only appears at late epochs is consistent with the idea that it is an effect that happens over many orbital periods.  \citet{ludlow19} pointed out that overly small softening values can actually decrease resolution owing to two-body scattering effects, so \simba's adaptive gravitational softening may exacerbate this issue.  It could also be that there is some missing physics in \simba\ that compacts low mass galaxies during quenching~\citep{tacchella_2017}, but given that \simba\ does produce quite compact galaxies at $z\sim 2$, we favor the explanation of numerical heating.  We note that star-forming galaxies will not suffer from this heating as much, because its stars were formed more recently in a thin disk. }

Looking at the higher redshifts, at $z=2$ the SF galaxies are significantly smaller than the observations, by a factor of \mycolor{$\sim 1.5$}. The passive galaxies are in good agreement with the observations at this redshift, however the SF galaxies represent the majority of the population. By $z = 1$, there is a larger population of passive galaxies which are in broad agreement with the observations, and SF galaxies are still smaller than their observational counterparts. The small sizes of the high redshift galaxies indicate that the \simba\ galaxies grow more rapidly since $z = 2$ than the real galaxies, suggesting that the growth modes for \simba\ galaxies differ from that in real galaxies.

\begin{figure}
	\includegraphics[width=\columnwidth]{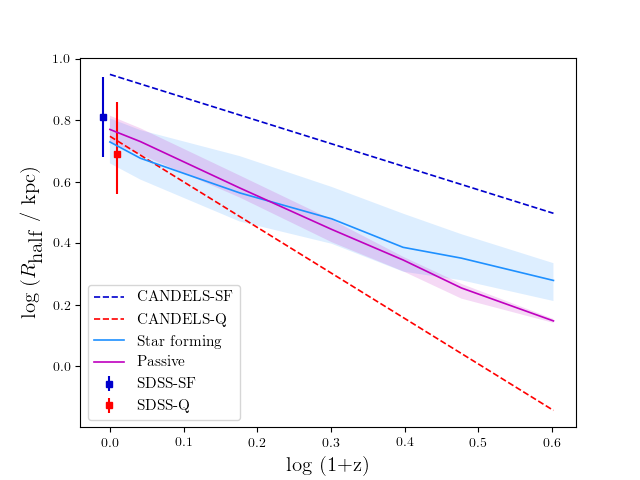}
	\vskip-0.1in
    \caption{Evolution of the V band half-light radii of star forming (light blue) and passive (magenta) central galaxies at $\mstar \sim 5 \times 10^{10} \msolar$. The solid lines show the running medians at each redshift and the shaded regions enclose 50\% of the data. The best fits to the evolution of the median sizes are \mycolor{$5.2 \, (1+z)^{-0.78}$ and $5.9 \, (1+z)^{-1.06}$} for star-forming and passive centrals respectively. The dashed lines show the corresponding redshift evolution for observations of V band galaxies sizes from \citet{van-der-wel_2014}, separated into star forming (blue) and quiescent (red) galaxies. The squares at $z=0$ are the corresponding R band half-light radii from SDSS \citep{zhang_2019}, offset by \(\pm\)0.01 for clarity.}
    \label{fig:redshift_medians_5e10}
\end{figure}

Figure \ref{fig:redshift_medians_5e10} quantifies the median size growth rate, showing the redshift evolution of $\rhalf$ in the V band for central galaxies with $\mstar$ within 5\% of \(5 \times 10^{10} M_{\odot}\). The galaxies are separated into star forming and passive using the same sSFR > \(10^{-1.8 + 0.3z}\)Gyr\(^{-1}\) cut as before. We choose this mass range to compare to the evolution of galaxies with \(5 \times 10^{10} M_{\odot}\) in \citet{van-der-wel_2014}, which are shown in the figure. We also show the \(z=0\) size measurements for this mass from SDSS \citet{zhang_2019}. By focusing on a particular mass, this plot allows us to examine the redshift dependence of the galaxy sizes.

The best fits for the evolution of the median sizes in \simba\ give sizes that scale as  \mycolor{\((1+z)^{-0.78}\) for the SF galaxies, consistent with observations showing a scaling of \((1+z)^{-0.75}\) for star-forming systems \citep{van-der-wel_2014}. For passive galaxies, the median sizes scale as \((1+z)^{-1.06}\), which is essentially a \((1+z)^{-1}\) evolution for the passive galaxy populations. This is consistent with expectations for a simple disk formation model~\citep{MoMaoWhite_98}, and that passive galaxies do not undergo any compaction when they quench out of the star-forming sequence. However, this is inconsistent with observations showing a scaling of  \((1+z)^{-1.48}\) for passive systems \citep{van-der-wel_2014}. } We note that our star-forming galaxy amplitude appears too low when compared to \citet{van-der-wel_2014}, but their fitting function at $z=0$ also lies noticeably above the SDSS measurement from \citet{zhang_2019}; it is beyond the scope here to examine why these two observational results disagree, albeit mildly. Finally, we note that our spatial resolution is fixed in comoving coordinates, which means that it scales as \((1+z)^{-1}\) in physical coordinates. Hence \mycolor{the scalings of passive galaxies} are consistent with an evolution in the numerical softening length, although our actual softening values are nominally smaller than galaxy sizes.

\mycolor{Interestingly, including dust extinction in our computation of $\rhalf$ has a substantial effect on the star forming galaxies. Without dust extinction, the size evolution goes as $(1+z)^{-1.16}$ and $(1+z)^{-1.11}$ for star forming and passive galaxies, respectively, making the evolution of star forming and passive galaxy evolution essentially the same. The substantial change in size evolution of the star forming galaxies is due to the increase in size at high redshift due to dust. Dust attenuation obscures light preferentially at the centres of galaxies, increasing the sizes, thus bringing the sizes into closer agreement with observations than without dust. This is particularly true for high redshift star forming galaxies as these objects contain the most dust \citep{li_2019}, while for passive galaxies with little dust the effect is weak.}

Other simulation projects have had varying levels of success in reproducing the galaxy sizes.
Illustris TNG is able to reproduce the mass-size relation of both SF and quenched galaxies at \(z=0\)~\citep{genel_2018}, showing good agreement with SDSS \citep{shen_2003, bernardi_2014} and \citet{van-der-wel_2014} extrapolated to \(z=0\). They are able to do this in part because they tune their simulation to match the \(z=0\) mass-size relation. However, the success of having quenched galaxies smaller than star-forming is something \simba\ fails, potentially in part because TNG has $\approx 20\times$ better mass resolution than \simba.
Likewise, EAGLE \citep{crain_2015, schaye_2015} has demonstrated that they match the low redshift \citet{shen_2003} SDSS measurements of star-forming galaxy sizes, though once again they tune their simulation to do so.  They are also able to get quenched galaxies smaller than star-forming, with a mass resolution $\approx 10\times$ better than \simba's.  
Horizon-AGN \citep{dubois_2014} has shown that their disk-dominated galaxy sizes are in agreement with \citet{van-der-wel_2014} at \(z=0.25\) to within a factor of \(\sim\) 2, but similar to \simba their elliptical galaxies are less compact than their disk galaxies \citep{dubois_2016}. They attribute this discrepancy to their limited spatial resolution, which is comparable to the resolution in \simba.

In summary, \simba\ produces low-$z$ SF galaxy half-light sizes in good agreement with observations. These constitute the galaxies we are most concerned with for the profiles in this work.  However, predicts that quenched galaxies have slightly larger sizes than star-forming systems at $z=0$ which is opposite to what is observed.  For the rest of this paper, we will investigate the radial profiles of star-forming or GV galaxies in various physical quantities, scaled by the half-light radii, primarily at $z=0$. We will not consider galaxies that are fully quenched.  Hence while it is a notable discrepancy that \simba\ does not reproduce the sizes of today's quenched galaxies, this is not critical for the results in the remainder of this work.

\section{Radial Profiles}\label{sec:radial}

\subsection{Galaxy selection}\label{sec:belfiore_selection}

We now examine galaxy radial profiles in \simba, focusing mainly on star-forming and GV systems at $z=0$, though we will look at redshift evolution in \S\ref{sec:zevol}.  We will focus on massive galaxies with $\mstar \geq 10^{10} \msolar$, corresponding to $\ga 550$ star particles, in order to ensure we can get sufficient resolution for robust profile measurements, and also because this is where observations for comparison are most abundant and secure.  We separate star-forming and GV galaxies via a cut in SFR(\mstar), described below.  We will not consider quenched galaxies further.

\begin{figure}
	\includegraphics[width=\columnwidth]{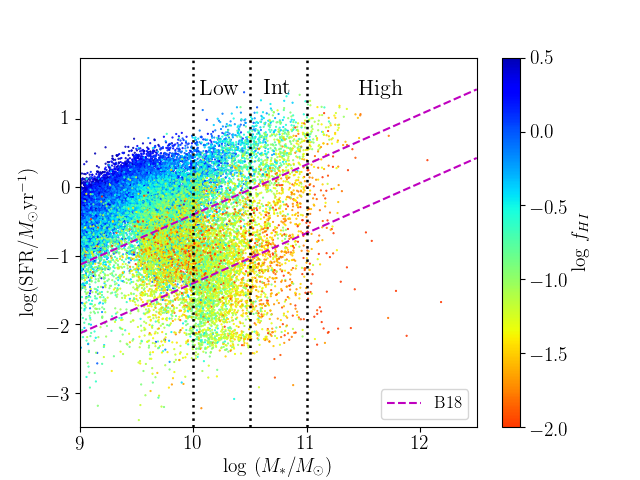}
    \caption{Star formation rate--stellar mass relation for all galaxies in the \simba 100 \(\hmpc\) box at $z=0$, colour coded by their \HI\ mass to stellar mass ratio, \(f_{\HI}\). The magenta dashed lines show the selection of the GV galaxies from \citet{belfiore_2018}, with the upper and lower lines corresponding to their SFMS and SFMS minus 1 dex, respectively. The black dotted lines show the three considered stellar mass bins labelled low ($10^{10}<\mstar<10^{10.5}M_\odot$), intermediate ($10^{10.5}<\mstar<10^{11}M_\odot$) and high mass ($\mstar>10^{11}M_\odot$).}
    \label{fig:b18_sample}
\end{figure}

Figure~\ref{fig:b18_sample} illustrates the way we select our simulated galaxy sample for this work. This shows the star-forming main sequence, \mstar vs. SFR, for all galaxies with $\mstar > 10^{9} \msolar$ in the 100\(h^{-1}\) Mpc \simba\ box at \(z = 0\). Points are colour coded by their \HI\ mass to stellar mass ratio, $\fHI$.  Vertical dotted lines denote the low, intermediate, and high mass bins that will be used for this work; all galaxies above $\mstar > 10^{11} \msolar$ are grouped into the high mass bin. We will not consider galaxies with $\mstar < 10^{10} \msolar$ further. The magenta lines demarcate the GV, as we discuss below.

This figure shows the usual structure of the main sequence. There is a blue cloud of star-forming galaxies extending towards low masses.  As \mstar increases there are more quenched galaxies with low SFR, primarily as a result of AGN jet feedback~\citep{dave_2019}, resulting in the development of a red sequence. As galaxies quench from the SF blue cloud to the red sequence, they move across \mstar--SFR space in the transitional GV region. There are occasional rejuvenations in the other direction, but they are quite rare in \simba, and additionally galaxies do not tend to loiter in the GV. It turns out the timescales to quench are bimodal, with slow quenching times of $0.1 \times t_{\textrm{Hubble}}$ and fast quenching times of $0.01 \times t_{\textrm{Hubble}}$ \citep[see][for a full analysis of quenching times and rejuvenations in \simba]{rodriguez-montero_2019}. 
Thus in \simba the vast majority of GV galaxies are on their way to being quenched eventually.

To demarcate the GV, we follow the definition of \cite{belfiore_2018}. Their star forming main sequence (SFMS)  is given by: 
\begin{equation}
    \log (\textrm{SFR}/\msolar\ \textrm{yr}^{-1}) = 0.73\ \log(\mstar/\msolar) - 7.33,
\end{equation}
with a scatter of 0.39~dex. The upper magenta dashed line in Figure \ref{fig:b18_sample} is the lower boundary of their star SF galaxies which corresponds to $1\sigma$ below the SFMS. They define GV galaxies as those with with SFR down to 1~dex below this, which is indicated by the lower magenta line.  These demarcations have a sub-unity slope, so they do not directly correspond to fixed cuts in sSFR, but for our mass range, star-forming galaxies defined this way typically have sSFR$\ga 10^{-10.5}$yr$^{-1}$.  

\begin{table*}
    \centering
    \caption{Median global galaxy properties \mycolor{and number of galaxies} for the three mass bins for SF and GV galaxies.}
    \begin{tabular}{c|c c c | c c c }
        \hline
        Median & & Star forming & & & GV &  \\ 
        &  Low & Int & High & Low & Int & High\\
        \hline
        \mycolor{$N_{\textrm{gal}}$} & \mycolor{1767} & \mycolor{603} & \mycolor{105} & \mycolor{1465} & \mycolor{373} & \mycolor{53} \\
        $\textrm{log} (\mstar / \textrm{M}_{\odot})$ & 10.2 & 10.7 & 11.1 & 10.2 & 10.7 &  11.1\\
        SFR (M$_{\odot} \textrm{yr}^{-1}$) & 1.56 & 3.64 & 6.18 & 0.14 & 0.39 & 1.03\\
        log (sSFR / yr$^{-1}$) & -9.98 & -10.16 & -10.36 & -11.04 & -11.12 & -11.14\\
        $\textrm{log} (M_\HI / \textrm{M}_{\odot})$ & 9.8 & 9.9 & 9.9 & 9.1 & 9.3 & 9.4 \\
        $\textrm{log} (M_{\textrm{H}_2} / \textrm{M}_{\odot})$ & 9.7 & 10.1 & 10.3 & 9.1 & 9.4&  9.6\\
        \mycolor{$\rhalf$ (kpc)} & \mycolor{3.8}& \mycolor{5.4} & \mycolor{6.8} & \mycolor{4.2} & \mycolor{5.2} &  \mycolor{6.0}\\
        \hline
    \end{tabular}
    
    \label{tab:median_data}
\end{table*}

With this selection, in \simba\ at $z=0$ we obtain 1767 star-forming galaxies in the low-mass bin ($10^{10}\msolar <\mstar<10^{10.5}\msolar$), 603 in the intermediate-mass bin ($10^{10.5}\msolar <\mstar<10^{11}\msolar$), and 105 in the high-mass bin ($\mstar>10^{11}\msolar$). For GV galaxies, these mass bins contain 1465, 373, and 53 galaxies, respectively.
Table~\ref{tab:median_data} shows median values for various galaxy properties in our star-forming and GV samples.

\subsection{Star-forming vs GV profiles}

We now examine the profiles in various physical quantities of the star-forming and GV samples defined as above. To generate profiles, we first rotate each galaxy such that it is face on, aligned with the angular momentum vector of all its cold gas and stars. We compute individual SFR and \mstar surface density radial profiles from the gas and star particles, and use these to compute an sSFR profile for each galaxy:
\begin{equation}
    \textrm{sSFR} (R) = \frac{\Sigma_{\textrm{SFR}}(R)}{\Sigma_{\mstar}(R)},
\end{equation}
where the $\Sigma$ represents the surface density within an annulus centreed at radius $R$ in the subscripted quantity. Where radial bins contain no gas, we take the SFR to be zero. Changes in sSFR can be due to a change in molecular gas fraction (\(f_{\textrm{H}_2}\)) or the star formation efficiency (\textrm{SFE}):
\begin{equation}
    \textrm{sSFR} = \frac{\textrm{SFR}}{M_{\textrm{H}_2}} \frac{M_{\textrm{H}_2}}{\mstar} = \textrm{SFE} \times f_{\textrm{H}_2}.
\end{equation}
To isolate which of these is responsible for any trends in sSFR, we decompose our sSFR profiles into profiles of \(f_{\textrm{H}_2}\) and SFE. These profiles are computed from the profiles of SFR, \mstar and \(H_2\) surface density:
\begin{equation}
    \textrm{SFE}(R)  = \frac{\Sigma_{\textrm{SFR}}(R)}{\Sigma_{\textrm{H2}}(R)}, \qquad f_{\textrm{H}_2}(R) = \frac{\Sigma_{\textrm{H2}}(R)}{\Sigma_{\mstar}(R)}. 
\end{equation} 
We centre profiles on the position of the galaxy's central black hole (which in \simba\ is tied to the location of the lowest potential), or in rare cases where there is no black hole then we choose the centre of mass of the star particles; our results are essentially unchanged if we always just use the centre of mass.
All profiles are normalised to the half-light radius $\rhalf$, that we compute as in \S\ref{sec:sizemass} using the unextincted SDSS $r$ band light at $z=0$ for comparison to low-$z$ SDSS data, and the unextincted rest-frame $V$ band for higher redshift comparisons, in order to mimic the band typically quoted from observations.  Our results are not sensitive to this choice within rest-frame optical bands.  

\begin{figure*}
	\includegraphics[width=\textwidth]{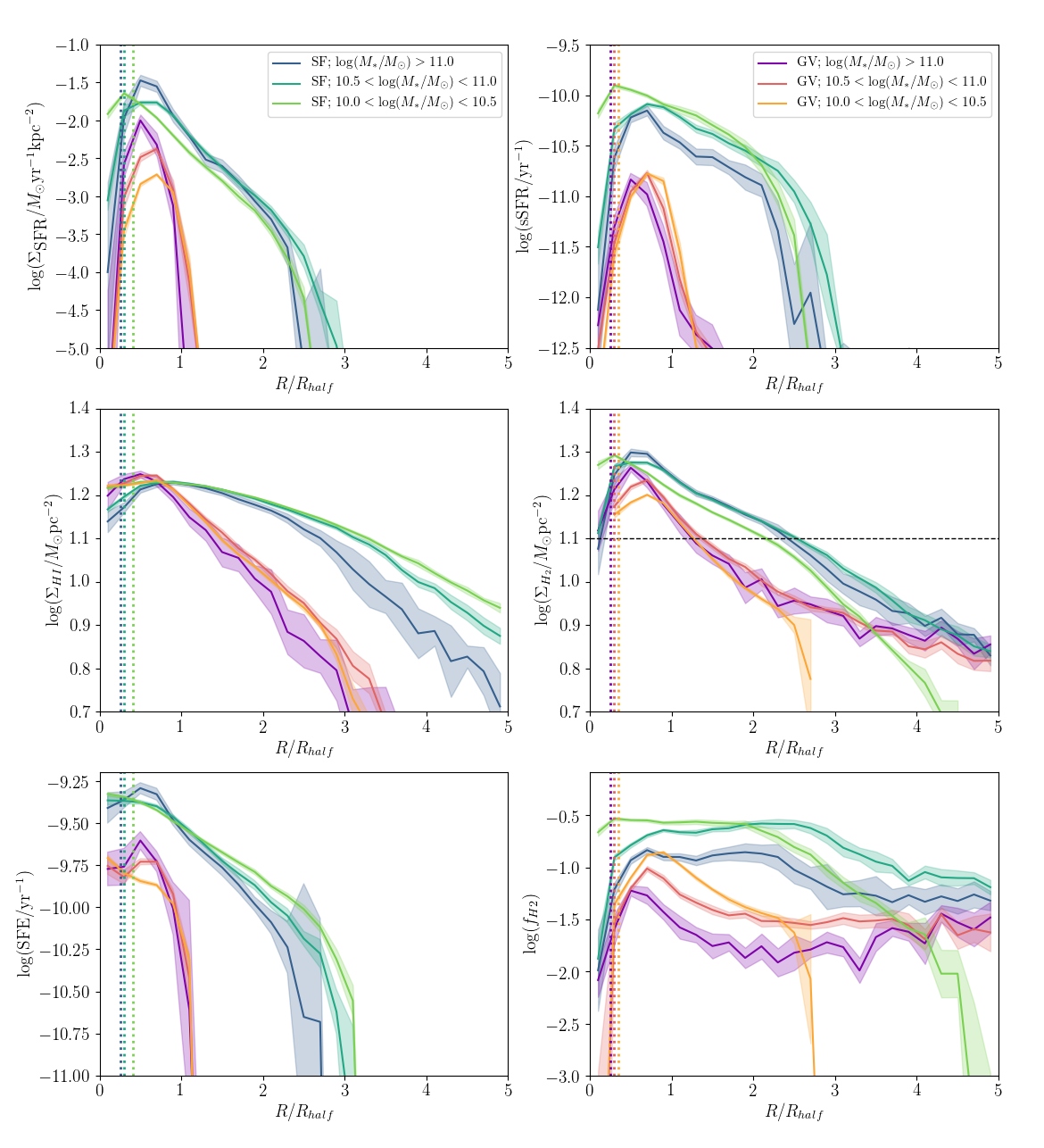}
	\caption{Radial profiles of SFR, sSFR, \HI, \(\rm{H}_2\), SFE and \(f_{H2}\) as a function of stellar mass. The green/blue lines show the SF galaxy profiles, and the orange/purple lines show the GV galaxy profiles, in low, intermediate, and high mass bins as indicated. The displayed radial profiles are Tukey biweights of the individual galaxy profiles in each mass bin. The light shaded regions around each line show the Tukey biweight scale estimator divided by \(\sqrt{N}\), combined with cosmic variance uncertainties from jackknife resampling over the 8 simulation sub-octants. \mycolor{The vertical dotted lines show the spatial resolution in units of the median half-light radius for each mass bin, computed separately for SF and GV galaxies.} Overall, GV galaxies show substantially different profiles than SF galaxies, with globally lower star formation and gas content, with the star formation confined to within $\rhalf$ as opposed to out to several half-light radii, and with a strong central dip in the SFR and sSFR profiles. This points towards two distinct mechanisms: inside-out quenching causing the suppression of star-forming gas in the central regions, and outside-in quenching suppressing the star formation efficiency in the outskirts.     
    }
    \label{fig:sf_profiles}
\end{figure*}

Figure \ref{fig:sf_profiles} shows radial surface density profiles scaled by $\rhalf$ for SFR, sSFR, \HI, \(\rm{H}_2\), SFE and \(f_{H2}\), separated into the mass bins shown in Figure \ref{fig:b18_sample}. The blue$\to$green lines show the SF galaxies; the purple$\to$orange lines show the GV galaxies.  In each mass bin, the overall profile is the Tukey biweight of the individual profiles. The Tukey biweight is a robust estimator for the mean that ignores outlying points~\citep[see e.g.][]{belfiore_2018, starkenburg_2019}. It is qualitatively similar to a median and also gives a robust estimation of the sample standard deviation (referred to as the biweight scale estimator). We also compute the error coming from cosmic variance via jackknife resampling over the 8 simulation sub-octants.  We add this error in quadrature with the Tukey biweight scale estimator divided by $\sqrt{N}$ (where $N$ is the number of galaxies in the mass bin), to obtain the shaded region around each line. \mycolor{The vertical dotted lines show the spatial resolution in units of the median $\rhalf$ for each mass bin, computed separately for SF and GV galaxies. Predictions on scales smaller than these lines are likely to be compromised by resolution effects.}

These profiles illustrate the stark structural differences between SF and GV galaxies in their star formation and gas profiles.  Overall, the GV galaxies have lower star formation and cold gas content at almost every radius compared to their SF counterpart at same mass. This is not unexpected, since these galaxies are on their way to being quenched, and thus will end up with very low SF and cold gas contents. For both SF and GV galaxies, the total star formation rate increases with stellar mass for both galaxy types, as quantified in Table~\ref{tab:median_data}. The SFR profiles show that star formation activity drops at a similar rate for all masses beyond $R\ga 0.5\rhalf$, but within this radius the profiles are mass-dependent.  For the SF galaxies, the highest mass galaxies show the strongest reduction of SFR in their centres, whereas for GV galaxies SFR drops towards zero in the centre for all mass bins similarly. The SF galaxies show more extended star formation, dropping to near zero SFR around $3\rhalf$, as opposed to $1.25\rhalf$ for the GV galaxies.

The sSFR profiles (upper right panel) show the same trend as the SFR; in general the galaxies have undergone quenching at both their centres and their outskirts, with a band of star formation occurring between $\sim 0.25-1\rhalf$ for the GV galaxies and between $\sim 0.5-2.5\rhalf$ for the SF galaxies. A decrease in sSFR may be due to either an increase in stellar mass or a decrease in SFR~\citep{spindler_2018}; the suppression of the SFR shows that the sSFR suppression is not simply due to the large stellar mass of the central galactic bulge. There is also a mass dependence -- a higher stellar mass translates to a lower overall level of sSFR, as expected from the sub-linear slope of the star-forming main sequence. The highest mass galaxies generally have the largest black holes and the most powerful AGN feedback, so they are expected to be in the process of quenching.  Indeed, high-mass SF galaxies also show some sSFR (and SFR) suppression in the centre, indicating that these galaxies are likely affected by the same mechanism(s) as the GV galaxies. This suggests that massive SF galaxies are in the early stages of the quenching process that has more substantially affected the GV galaxies, which is consistent with the idea of slow transition to the red sequence in massive SF galaxies inferred from observations~\citep{schawinski_2014}.

The middle panels of Figure~\ref{fig:sf_profiles} show the cold gas (\HI\ and H$_2$) mass density profiles.  For star-forming galaxies, similar to the \(\Sigma_{SFR}\) profiles, the \(\Sigma_{\textrm{HI}}\) profiles show that \HI\ and H$_2$ surface densities decrease with increasing radial distance, but are suppressed in the centre of the galaxy. There is only a weak dependence on mass, as the \HI\ profile is fairly universal with the only difference being a somewhat more rapid drop in the outskirts in more massive galaxies, and the H$_2$ profile being mildly lower for the lowest mass galaxies. Observationally, it has been noted that SF galaxies exhibit a near-universal cold gas surface density profile when scaled by size~\citep{bigiel_2012}, which is qualitatively consistent with what we find.  Essentially, the increase in overall galaxy size yielding a larger gas content is offset by the decrease in cold gas content in more massive galaxies, resulting in profiles that are broadly independent of mass.

Comparing the SF galaxies to the GV galaxies shows marked differences.  In \HI, the gas content has dropped sharply in the outskirts relative to SF galaxies.  This could owe to starvation not replenishing cold gas in the outskirts as it moves inwards to form stars, evaporation by a growing hot halo that is more prevalent around GV galaxies than SF systems at the same mass, and/or environmental processes where interactions with nearby satellite galaxies have dynamically heated the cold gas. We will examine satellite vs. central profiles in \S\ref{sec:cens_sats}.

For H$_2$, the situation is more curious.  In SF galaxies, the H$_2$ surface density peaks at a higher value, shows a greater drop in the middle, and drops off more quickly in the outskirts than the \HI.  All these trends are consistent with a higher H$_2$ fraction in denser gas, with the exception of the central dip; in this case, it could be that extraplanar gas in the foreground of our face-on galaxies is predominantly in \HI\ form, and so fills in the central region in projection.

For the GV galaxies, there is a drop in the H$_2$ content in the inner parts relative to SF galaxies, but in the outskirts, the H$_2$ profile is actually shallower in the outskirts, particularly for more massive galaxies. Yet despite the H$_2$ having quite an extended profile, the SFR surface density remains well confined to the central region.  In other words, there is still substantial H$_2$ in the outskirts of GV galaxies, but it is not forming stars.  This could owe to the fact that the physical densities are substantially lower in the outskirts, and since in our simulations SFR$\propto \rho^{1.5}$, this can cause a strong drop in SFR even if the projected $\Sigma_{\textrm{H}2}$ remains high.  Interestingly, if one postulates a threshold H$_2$ surface density of $\log\Sigma_{\textrm{H}2}=1.1$ in order to have sufficient physical density for star formation (dotted horizontal line), then this would truncate star formation at $\sim 3\rhalf$ in SF galaxies and $\sim\rhalf$ in GV galaxies, which is essentially what is seen in the $\Sigma_{SFR}$ plot in the upper right. Such a surface density threshold is approximately coincident with the turn-down in $\Sigma_{SFR}$ seen in the K-S at low gas surface densities (Fig. \ref{fig:kennicutt}).

The bottom two panels of Figure~\ref{fig:sf_profiles} quantify the connection between star formation and dense gas more clearly.  Since sSFR$=$SFE$\times f_{H2}$, we can subdivide the sSFR profile into profiles for SFE (left) and $f_{H2}$ (right) to better understand why GV galaxies have suppressed star formation.  In essence, the plot in the upper right is convolution of the two bottom plots.

It is clear that the H$_2$ fractions (bottom right), while lower for GV galaxies, have a similar radial trend between the SF and GV galaxies: the gas fraction is relatively flat except in the central region where it drops.  Meanwhile, the SFE shows a rapid decline with radius, and no central drop.  We thus see that the there are two separate effects which conspire to take the sSFR profiles from SF to GV: In the outer regions, the sSFR is suppressed owing to a rapid SFE decline in the outskirts; this is primarily governed by the physical density of the star-forming gas. In contrast, in the innermost region the H$_2$ fraction drops quickly, and hence the central hole in sSFR in GV and massive SF galaxies primarily owes to molecular gas being removed either by heating or expulsion.  Gas in the centre is forming stars at a similar efficiency as gas at the peak in the SFR profile, but there is simply much less of it.

In summary, GV galaxies show substantially different profiles than SF galaxies, with overall lower star formation and gas content, and the star formation being confined to within $\rhalf$ as opposed to out to several half-light radii.  There is a strong central dip in the SFR as well as sSFR profiles, which occurs in all GV galaxies along with massive SF galaxies. The SFR drop in GV systems thus appears to be driven by two different effects in the inner and outer regions. In the central region, the amount of star-forming gas is suppressed, while in the outskirts, molecular gas is still present but has a suppressed efficiency of forming into stars.  Thus it appears that quenching in \simba\ galaxies occurs both inside-out and outside-in, and may indicate two distinct physical mechanisms.  We will examine which AGN feedback mechanisms are responsible for these effects in \S\ref{sec:feedback}.  Next we conduct a more careful comparison to observations of SF vs. GV galaxy profiles.

\subsection{Comparison to observations}

\begin{figure*}
	\includegraphics[width=\textwidth]{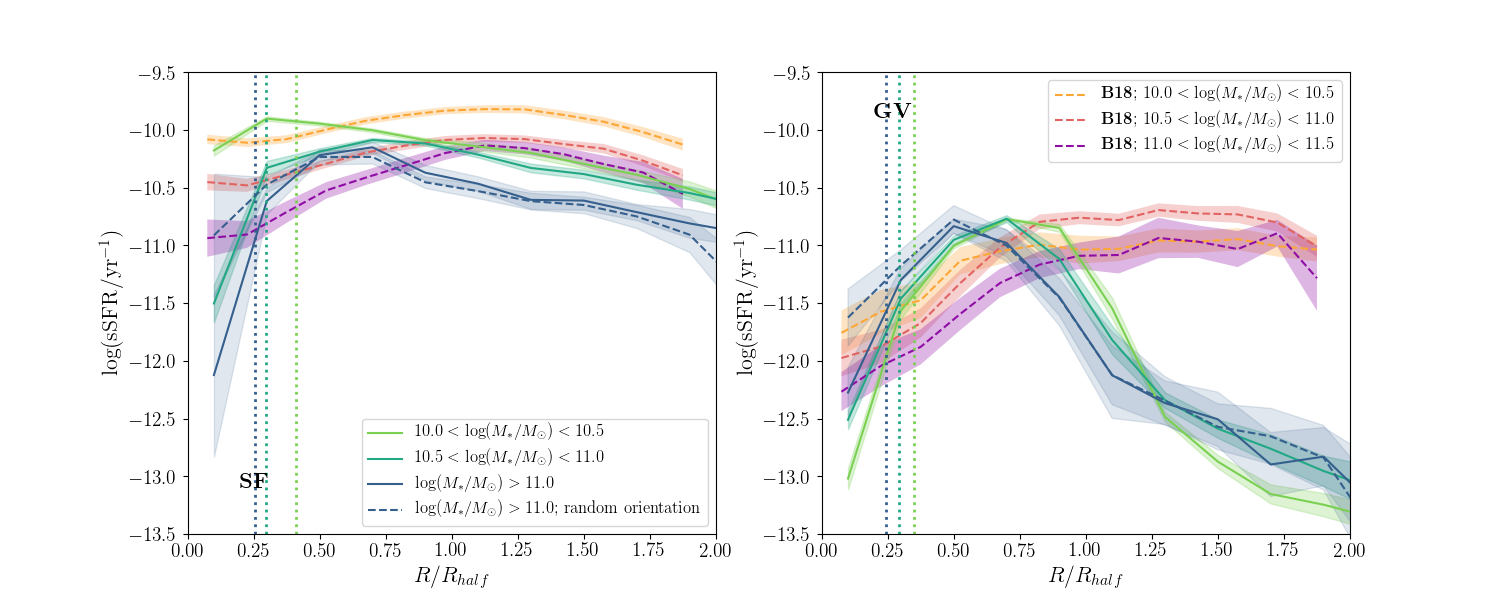}
	\vskip-0.2in
    \caption{The sSFR radial profiles for all SF (left) and GV (right) galaxies for the full \simba\ run, with increasing \mstar bins shown in green to blue. The solid lines show the profiles in the frame rotated such that the galaxies are face-on, the dashed line for the highest mass bin shows the non-rotated profiles (i.e. the randomly orientated case). Observations of sSFR profiles of galaxies from MaNGA SDSS \citep{belfiore_2018} are shown as the pink/purple lines. The radial profiles are Tukey biweights of the individual galaxy profiles in each mass bin. The light shaded regions correspond to the Tukey biweight scale estimator divided by \(\sqrt{N}\), combined with cosmic variance uncertainties from jackknife resampling over the 8 simulation sub-octants.  \mycolor{The vertical dotted lines show the spatial resolution in units of the median half-light radius for each mass bin, computed separately for SF and GV galaxies.} SF galaxies show a reasonable good agreement with the MaNGA data at low mass, but the inside-out quenching occurring in massive star-forming galaxies appears to be too strong. For the GV galaxies, \simba reproduces a drop in sSFR in the central regions seen in the data, however in the outskirts the sSFR is also suppressed, in conflict with the data.}
    \label{fig:ssfr_s50j7k_belfiore}
\end{figure*}

The sSFR profiles of star-forming versus GV galaxies has been measured in the SDSS MaNGA Survey by \citet{belfiore_2018}.  They showed that the sSFR profiles of GV galaxies tend to be strongly suppressed in the centres relative to SF galaxies. \citet{starkenburg_2019} examined these trends in the Illustris and EAGLE simulations, and surprisingly found that despite these models quenching galaxies globally as observed, the sSFR profiles of similarly-selected GV galaxies did not show any strong central suppression in either simulation, but rather a centrally concentrated sSFR profile that was qualitatively similar to that in SF galaxies.  They thus highlighted this comparison as a key test of how galaxies quench radially in models, one that some state of the art simulations fail to satisfy. \mycolor{In the more recent Illustris TNG simulation, radial profiles show some central suppression of star formation, but only in the most massive galaxies \citep{nelson_2019}.} In this section we undertake the \citet{starkenburg_2019} comparison in \simba. 

Figure \ref{fig:ssfr_s50j7k_belfiore} shows the comparison of sSFR profiles for SF galaxies (left panel) and GV galaxies (right panel) in \simba, with increasing \mstar bins shown in green to blue, versus profile in the same \mstar bins from \citet[orange to purple]{belfiore_2018}. 
All profiles are scaled to the $r$-band half-light radius $\rhalf$, and are computed over all profiles in each mass bin using the Tukey biweight estimator as done in \citet{belfiore_2018}. These \simba profiles are exactly as plotted in Figure~\ref{fig:sf_profiles}, but here we zoom in on the central region ($R/\rhalf<2$), split SF and GV into separate panels, and show observations overlaid.  

For the SF galaxies (left panel), \simba predicts sSFR profiles that are qualitatively consistent with observations. From the centre outwards, the profiles show a rise in the central region, and then a mostly flat profile in the outskirts.  The rise is faster in higher mass galaxies, indicating greater suppression of central SFR in these systems. However, there are some clear discrepancies versus data, particularly for higher-mass SF galaxies. First, the sSFR values peak at smaller radii in \simba ($R_{\rm peak}\sim 0.5\rhalf$) versus observations ($R_{\rm peak}\sim 1-1.5\rhalf$). A more blatant discrepancy is seen in the inner region, where the drop seen in the observations is not nearly as abrupt as that predicted in \simba for $\mstar >10^{10.5} \msolar$ galaxies. Hence the agreement between \simba and the MaNGA data is reasonable for lower-mass SF galaxies, but the inside-out quenching already occurring in massive star-forming galaxies appears to be too severe.

For the GV galaxies (right panel), it is clear that \simba produces a drop in the central sSFR.  This is in good agreement with the \citet{belfiore_2018} data, at least better than other current simulations~\citep{starkenburg_2019}. The sSFR starts at similar values at its peak in \simba\ and in the data (sSFR$\approx 10^{-11}$yr$^{-1}$), and drops by an order of magnitude or more towards the middle.  The main difference is that the decline is more gradual in the data, starting at around $\sim\rhalf$, while in the simulations it begins dropping inside $\sim 0.5\rhalf$.  We will discuss this further in \S\ref{sec:feedback}.

Now examining the outer parts of the sSFR profile ($R/\rhalf\ga 1$), we see that the sSFR in GV galaxies is also suppressed relative to SF galaxies in \simba\ at all masses in the outskirts.  This is clearly in conflict with the data.  Interestingly, EAGLE and Illustris likewise produce sSFR profiles that drop rapidly more quickly than observed in GV galaxies, so while \simba\ yields a central hole in sSFR in better agreement with data than those simulations, in the outskirts \simba\ is similar to other simulations. \mycolor{The discrepancy in the outskirts could be due in part to the conversion of H$\alpha$ to SFR in \citet{belfiore_2018}, which could have some contribution from non-star forming diffuse ionised gas. The contribution of dust-scattering to the H$\alpha$ emission of diffuse ionised gas in simulations is estimated to be at most 50\% of the total emission \citep{ascasibar_2016}. Depending on how radially-dependent this contribution is, it could make a difference to the outskirts of the observed radial profiles.}

We note that the \citet{belfiore_2018} analysis computes radial profiles in elliptical apertures to account for inclination, \mycolor{using elliptical Petrosian effective radii $R_e$ and inclinations from the NASA-Sloan catalogue (NSA $v1\_0\_1$\footnote{https://www.sdss.org/dr13/manga/manga-target-selection/nsa/}, \citealt{blanton_2011}) to construct de-projected radial profiles with elliptical annuli of semimajor axis $0.15R_e$.} We have approximated this process by first making all our galaxies face-on before calculating profiles. These methods should be identical in the case of a perfectly thin disk, but in our simulations, the stellar disks are not particularly thin.  A similar procedure would likely blur out the central region, and thus potentially mitigate the differences with the SF population.  We demonstrate how large an effect this may have by showing non-rotated profiles for the highest mass bin in Figure \ref{fig:ssfr_s50j7k_belfiore} -- the rotated and non-rotated profiles show almost no difference beyond \(0.5 \rhalf\), but in the central region the rotated profiles have much lower sSFR. We see that de-projecting the profiles emphasises the central suppression in star formation. 

Also, it is worth noting that the \citet{belfiore_2018} data explicitly removes galaxies with Seyfert-like line ratios.  Seyferts are typically large star-forming disks with strong AGN activity.  We are not currently able to identify Seyferts via line ratios in our simulation, so we have not mimicked this selection. It may be possible that such galaxies would have SFR profiles that drop rapidly in the middle owing to the putative nuclear AGN feedback, which would make the profiles of SF galaxies drop more quickly in the centres.  Unfortunately, it is difficult to measure inner SFRs in Seyferts owing to AGN contamination, which is precisely why these were excluded in observations.  Seyferts make up a relatively small fraction ($\sim 10\%$) of the overall disk population, but may contribute more to the most massive bins.  Hence this may explain part of the massive SF galaxy discrepancy.

Recall that \simba\ has two separate effects going on to suppress star formation in GV galaxies (\S\ref{sec:radial}): In the inner parts, this owes to removal of star-forming gas which lowers the gas fraction, while in the outer parts it owes to a lower star-forming efficiency.  It appears that the physics driving the inner suppression is roughly consistent with observations for GV galaxies, but that driving the outer suppression via a drop in the SFE is not.  It also appears that the onset of the inside-out quenching occurs in massive star-forming galaxies in \simba\ is quite strong, whereas such galaxies in observations show only a mild central suppression. Hence while reproducing the central sSFR drop in GV galaxies is a qualitative success of \simba, there remain substantial discrepancies in galaxy sSFR profiles.  To explore the physical drivers of these various effects, we now examine which feedback mechanisms are responsible for these trends.

\subsection{Black hole feedback dependence}\label{sec:feedback}

\begin{figure*}
	\includegraphics[width=\textwidth]{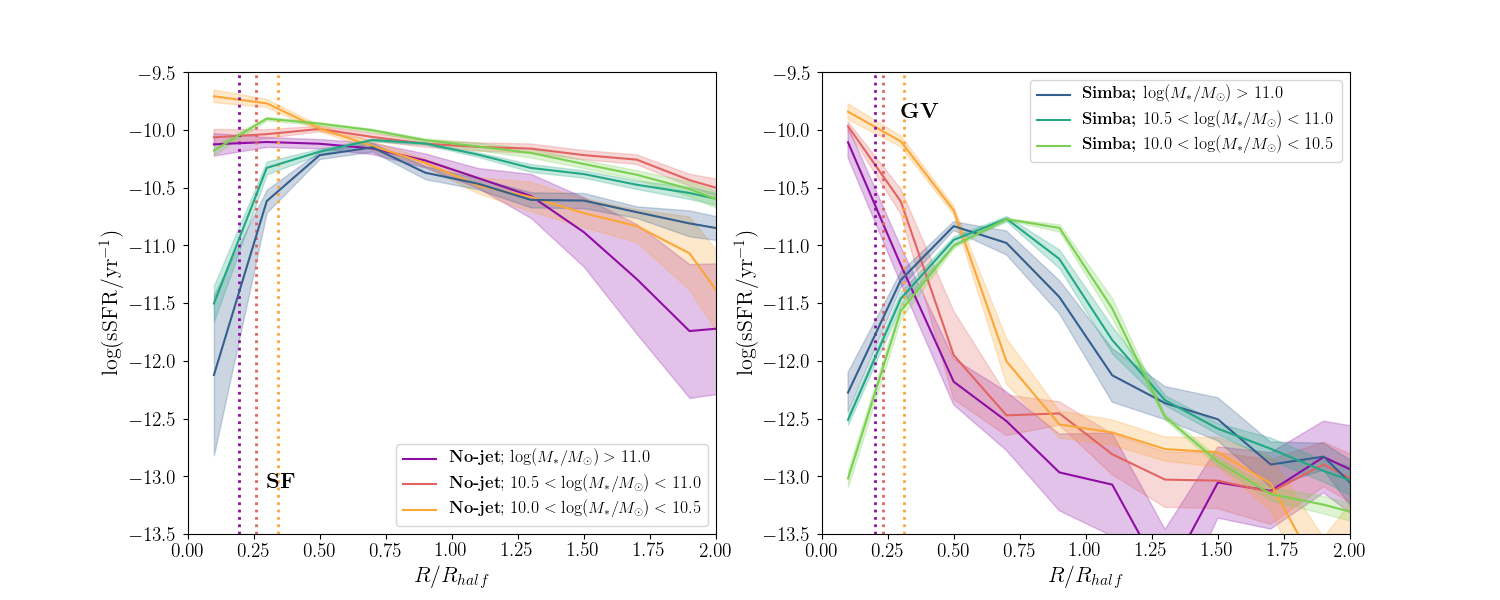}
	\vskip-0.1in
    \caption{The sSFR radial profiles for all SF (left) and GV (right) galaxies for the \simba without X-ray or jet feedback (pink/purple lines) and the full \simba (blue/green lines) $50\hmpc$ runs. The displayed radial profiles are Tukey biweights of the individual galaxy profiles in each mass bin. The light shaded regions around each line show the Tukey biweight scale estimator divided by \(\sqrt{N}\), combined with cosmic variance uncertainties from jackknife resampling over the 8 simulation sub-octants. \mycolor{The vertical dotted lines show the spatial resolution in units of the median half-light radius for each mass bin, computed separately for SF and GV galaxies.} The most notable difference in the sSFR profiles of SF galaxies occurs in the central region of more massive SF galaxies, where the {\it No-jet} run produces no dip, suggesting that it is a direct result of either jet or X-ray feedback. The differences for the GV galaxies are much more striking, showing a strongly centrally concentrated star formation, with the dropoff in the sSFR profile being stronger than in the full feedback case. Hence AGN feedback seems crucial for redistributing the star formation in GV galaxies at all radii.
    }
    \label{fig:ssfr_profile_nojet}
\end{figure*}

\begin{figure*}
	\includegraphics[width=\textwidth]{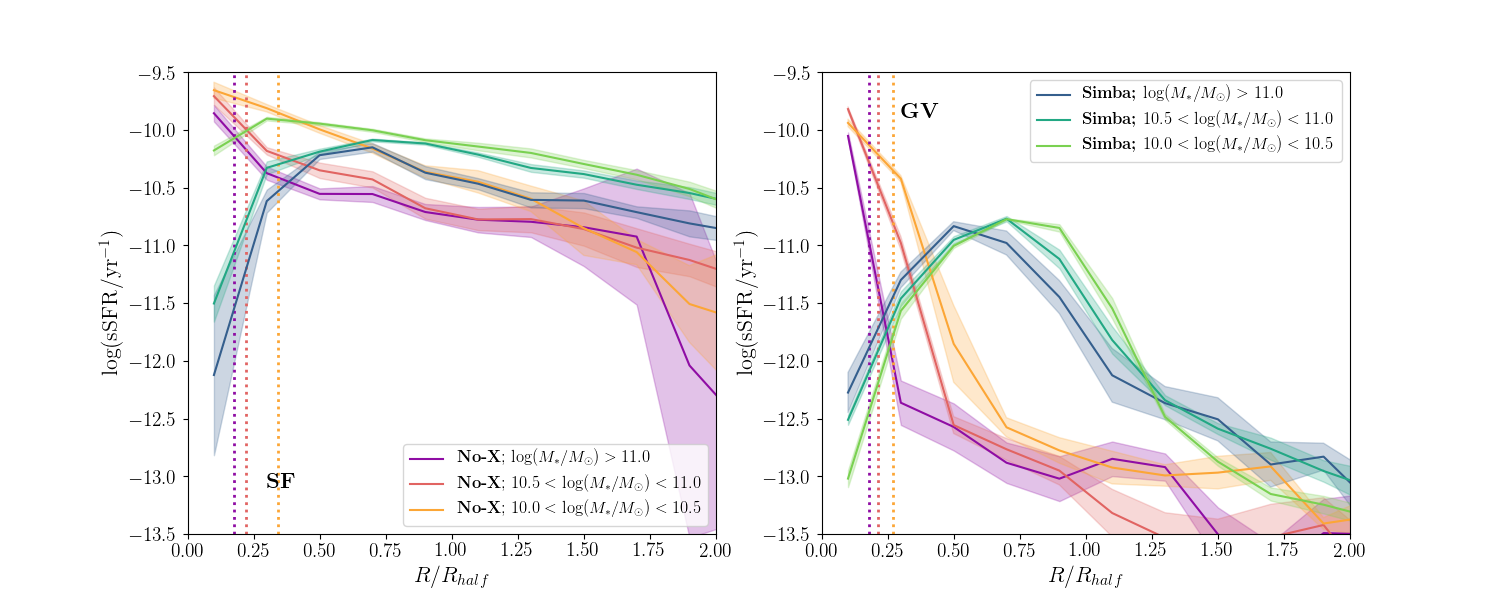}
	\vskip-0.1in
    \caption{The sSFR radial profiles for all SF (left) and GV (right) galaxies for the \simba without X-ray feedback (pink/purple lines) and the full \simba (blue/green lines) $50\hmpc$ runs. The displayed radial profiles are Tukey biweights of the individual galaxy profiles in each mass bin. The light shaded regions around each line show the Tukey biweight scale estimator divided by \(\sqrt{N}\), combined with cosmic variance uncertainties from jackknife resampling over the 8 simulation sub-octants. \mycolor{The vertical dotted lines show the spatial resolution in units of the median half-light radius for each mass bin, computed separately for SF and GV galaxies.} Turning on jets and leaving the X-rays off leads to even more centrally concentrated sSFR profiles that are suppressed in the outskirts for both SF and GV galaxies compared to the full \simba and {\it No-jet} runs. Jet feedback appears to have the overall effect of slightly suppressing star formation in the outskirts, causing that gas to move inwards in order to form stars in a more centrally concentrated manner. Hence jet feedback as implemented in \simba is not responsible for the central sSFR suppression observed by \citet{belfiore_2018}.}
    \label{fig:ssfr_profile_nox}
\end{figure*}

In \simba, AGN feedback is primarily responsible for quenching galaxies.  Of the three forms of AGN feedback implemented in \simba, the jet mode feedback is most directly responsible for quenching, the X-ray feedback by itself does not quench but is nonetheless important for fully quenching galaxies, and the radiative mode feedback is essentially irrelevant for quenching~\citep{dave_2019}.  The question is then, how do these various AGN feedback forms impact the profiles of quenching galaxies?  To answer this, here we compare our profiles in a full-physics \simba\ run versus two other runs: {\it No-jet}, where we turn off both X-rays and jets, and {\it No-X}, where we turn off only the X-ray feedback but leave jets on.  These are done using $50\hmpc$, $2\times 512^3$ particle runs, but the numerical resolution and input physics are otherwise identical to the full \simba\ run. 

Figure \ref{fig:ssfr_profile_nojet} shows the sSFR profiles for galaxies in the {\it No-jet} \simba\ run (i.e. without X-ray or jet mode feedback) shown in purple to orange. As in Figure~\ref{fig:ssfr_s50j7k_belfiore}, the left panel shows the SF galaxies, while the right panel shows the GV galaxies. We reproduce the results from the full \simba\ simulation for comparison (in blue to green), but leave off the \citet{belfiore_2018} observations to avoid confusion. Note that in the {\it No-jet} simulation, massive galaxies do not quench, but typically end up as star-forming or GV galaxies~\citep{dave_2019}.

The outer SF galaxy profiles are not markedly different over most radii with jet and X-ray feedback off versus in the full \simba\ model.  The sSFR values are only mildly higher in the No-jet case at all radii beyond $\ga 0.5\rhalf$. The profiles are also nearly identical for SF galaxies regardless of $\mstar$, whereas the full Simba model yields a stronger mass dependence, in better agreement with the observations.. The most notable difference occurs in the central region of more massive SF galaxies, where the {\it No-jet} run produces no sSFR dip in the central region. Hence we infer that this dip is a direct result of either jet or X-ray feedback. Although we do not show the observations from \citet{belfiore_2018} overlaid, the core sSFR is higher in the {\it No-jet} run compared to observations, particularly for the more massive galaxies.  Hence the \citet{belfiore_2018} data seems to require some suppression of core SF in massive galaxies, but not as much as in the full \simba\ run.  Outside the central region, the profiles are in good agreement with data.  This shows that AGN feedback has a modest but non-trivial impact on even star-forming galaxies, and at least some AGN feedback is already required to produce a central sSFR depression in massive SF systems.

Turning to the GV galaxies (right panel), the differences are much more striking. The {\it No-jet} sSFR profiles show strongly centrally concentrated star formation, with more compact extent (relative to $\rhalf$) for more massive systems.  Also, the GV profiles are quite a bit steeper than the SF galaxy profiles.  This shows that the reason these galaxies are in the GV is that star formation has been eroded in the outskirts, likely because these galaxies live in shock-heated hot halos~\citep{keres_2005} that is starving these systems of fresh gas.  In other words, without significant AGN feedback, suppression of star formation occurs outside-in.  This is exactly opposite to the way that GV galaxies are observed to be quenching, from the inside-out.

Interestingly, the {\it No-jet} GV profiles show an even stronger dropoff in the sSFR than in the full feedback case; thus if anything, AGN feedback appears to be puffing out the star forming gas relative to the stellar mass (or $r$-band light).  Hence it is not possible to solve the discrepancy between \simba\ and the \citet{belfiore_2018} data in the outskirts by simply saying that AGN should have no effect there.  Indeed, AGN appear crucial at all radii for redistributing the star formation in GV galaxies in a manner consistent with observations, for all galaxy masses probed here.

Given that the combination of jet and X-ray feedback appears to be crucial for altering the GV sSFR profiles to be in better qualitative agreement with data, it is interesting to ask which of these two modes is most responsible.  To examine this, we now examine {\it No-X} where we turn on the jet feedback, but leave the X-ray feedback off.

Figure~\ref{fig:ssfr_profile_nox} shows the resulting profiles from the {\it No-X} run, analogous to Figure \ref{fig:ssfr_profile_nojet}.  The {\it No-X} run does produce some quenched galaxies, but generally they do not have as low sSFR as observed.  Hence a histogram of sSFR's from this model does not agree with observations as it does for the full \simba\ run~\citep{dave_2019}.

Remarkably, turning on jets and leaving the X-rays off actually leads to even more concentrated sSFR profiles.  Now, even the SF galaxy profiles are clearly wrong -- they are centrally peaked, and are suppressed in the outskirts, relative to the full \simba\ and {\it No-jet} runs.  It appears that jet feedback has the overall effect of slightly suppressing star formation in the outskirts, causing that gas to move inwards in order to form stars in a more centrally concentrated fashion.  Recall that in \simba\ our jets are purely bipolar, and explicitly do not interact with surrounding gas until they are outside the ISM.  Hence it is not surprising that they do not suppress the central SF, but it is curious that they indirectly cause an enhancement, at least relative to the stellar mass distribution.

Moving to the GV galaxies (right panel), the central concentration of sSFR is now even more apparent than in the {\it No-jet} case.  In GV galaxies, the jet feedback is strongly suppressing the star formation in the outer regions leading to much steeper profiles relative to \simba, but the innermost sSFR is essentially unchanged from the SF galaxies in the left panel.  Clearly, jet feedback as implemented in \simba\ is not responsible for the central sSFR hole that is observed by \citet{belfiore_2018}.

It is only when we turn on X-ray feedback as in the full \simba\ run that we produce central suppression in the GV (as well as massive SF) population.  We conclude therefore that it is \simba's X-ray feedback that is responsible for creating the central depression in sSFR as observed by \citet{belfiore_2018} and others.

How does X-ray feedback as implemented in \simba\ cause inside-out quenching?  Our implementation of X-ray feedback represents a sub-resolution model for momentum input from the X-ray photons generated in the accretion disk.  In it, a kick is applied outwards from the black hole onto gas within the black hole accretion kernel, based on the X-ray radiative momentum input coupling to hydrogen gas as outlined in \citet{choi_2012}.  Since the momentum input drops as $1/r^2$ from the black hole, the gas closest to the black hole is most strongly kicked, which creates a hole in the cold gas and hence in the star formation.

Importantly, \simba\ implements a 2~kpc maximum radius for the black hole accretion kernel (or 256 nearest neighbours, whichever is smaller), and thus for the extent of direct X-ray feedback kicks.  In principle there is no reason why X-ray photon pressure should be limited to this radius; this was done purely for computational convenience.  Since there was already a neighboring particle list identified for the black hole accretion module out to (up to) 2~kpc, it was most straightforward to implement a kick on these pre-identified particles.  The inadvertent result of this is that X-ray feedback is only immediately felt out to $\la 2$~kpc.  This may explain why our full \simba\ profiles rise quickly out to a maximum at a $2-3$~kpc ($\sim 0.5\rhalf$); had we allowed our X-ray feedback to operate to larger radii, it is possible we would have generated a more gradual rise in sSFR out to larger radii, which would qualitatively be in better agreement with the observed profiles of \citet{belfiore_2018}.

The X-ray feedback does not fix the too-rapid dropoff in sSFR at $R\ga \rhalf$ in GV galaxies, seen at all masses and even with AGN feedback mostly off.  This remains something of an enigma.  This rapid dropoff also appears in GV galaxy profiles in EAGLE and Illustris~\citep{starkenburg_2019}, hence it seems to be a fairly generic outcome of current galaxy formation models: galaxies that have depressed overall star formation tend to have it particularly depressed in their outskirts, in clear disagreement with observations.  One commonality between all these simulations is that their subgrid AGN feedback models quench galaxies primarily by keeping the surrounding circum-galactic gas hot~\citep[e.g.][]{DaviesCrain2020}, albeit via differing mechanisms.  This is long known to be a successful approach to quenching~\citep{croton_2006,bower_2006,gabor_2015,dave_2016}. It is possible, however, that such preventive feedback preferentially suppresses star formation in the outskirts of galaxies by shutting off the accretion that would otherwise replenish an extended cold gas reservoir.  This appears to be in contradiction with observations.  One perhaps relevant point in \simba\ is that there is significant molecular gas in the outskirts, but it has low star formation efficiency~(Fig.~\ref{fig:sf_profiles}, bottom panels).  Adjusting the star formation prescription to have that gas continue to form stars would yield better agreement for sSFR profiles in the outskirts.  We leave such explorations for future work.  In any case, it appears to be a significant challenge for models to quench galaxies as observed globally, while retaining active star formation out to several half-light radii in transitional GV galaxies.

To summarise, \simba\ shows low-mass star-forming galaxy sSFR profiles and GV sSFR central depressions that are broadly in agreement with observations.  The central depressions owe specifically to X-ray feedback as implemented in \simba, which imparts outwards momentum to the gas surrounding the black hole.  Other simulations such as EAGLE and Illustris, which do not have such a mechanism, fail to match this. It is possible that Illustris-TNG may fare better, because although they do not implement X-ray feedback as in \simba, they tend to sphericalise the jet energy input by randomly re-orienting the jet at every timestep, and they do not shut off hydrodynamic interactions between the jet and the ISM as in \simba~\citep{weinberger_2018}. This could result in a qualitatively similar outward momentum injection.  One could envision that simply heating the ISM near the black hole might also be sufficient to create an sSFR hole, but this is essentially how the EAGLE AGN feedback model operates, and it does not succeed.  More generally, our results imply that current observations require some mechanism that evacuates gas from the central regions of galaxies during the quenching process, in a manner that operates approximately like \simba's X-ray feedback module.  This inside-out quenching, along with the unresolved discrepancies in the outskirts of GV sSFR profiles, represent key constraints on quenching prescriptions in current galaxy formation models.

\subsection{Satellites vs. centrals}
\label{sec:cens_sats}

So far, we have not distinguished centrals versus satellites, and simply considered all galaxies within our specified cuts.  However, satellites can experience environmental quenching processes that could in principle impact their profiles differently than internal processes such as AGN feedback.  
For instance, ram pressure and tidal stripping might remove gas preferentially from the outskirts, which would result in more compressed \HI\ profiles as seen in GV galaxies, but for reasons that do not involve AGN feedback.  Alternatively, they could be starved of gas infall owing to living within a hot halo, and thus have their SFR suppressed. \citet{spindler_2018} examined satellite vs. central profiles in MaNGA and found that satellites have overall lower sSFR at most radii vs mass-matched centrals, but that the suppression in the central region is similar.  They interpret this to suggest that satellites have lower sSFR overall owing to strangulation that cuts off their broader gas supply, but that the core sSFR suppression is a separate internal process.  Here we examine the profiles of satellite versus central galaxies in \simba\ to better understand how they are impacted by satellite-specific processes.

\begin{figure*}
	\vskip-0.1in
	\includegraphics[width=\textwidth]{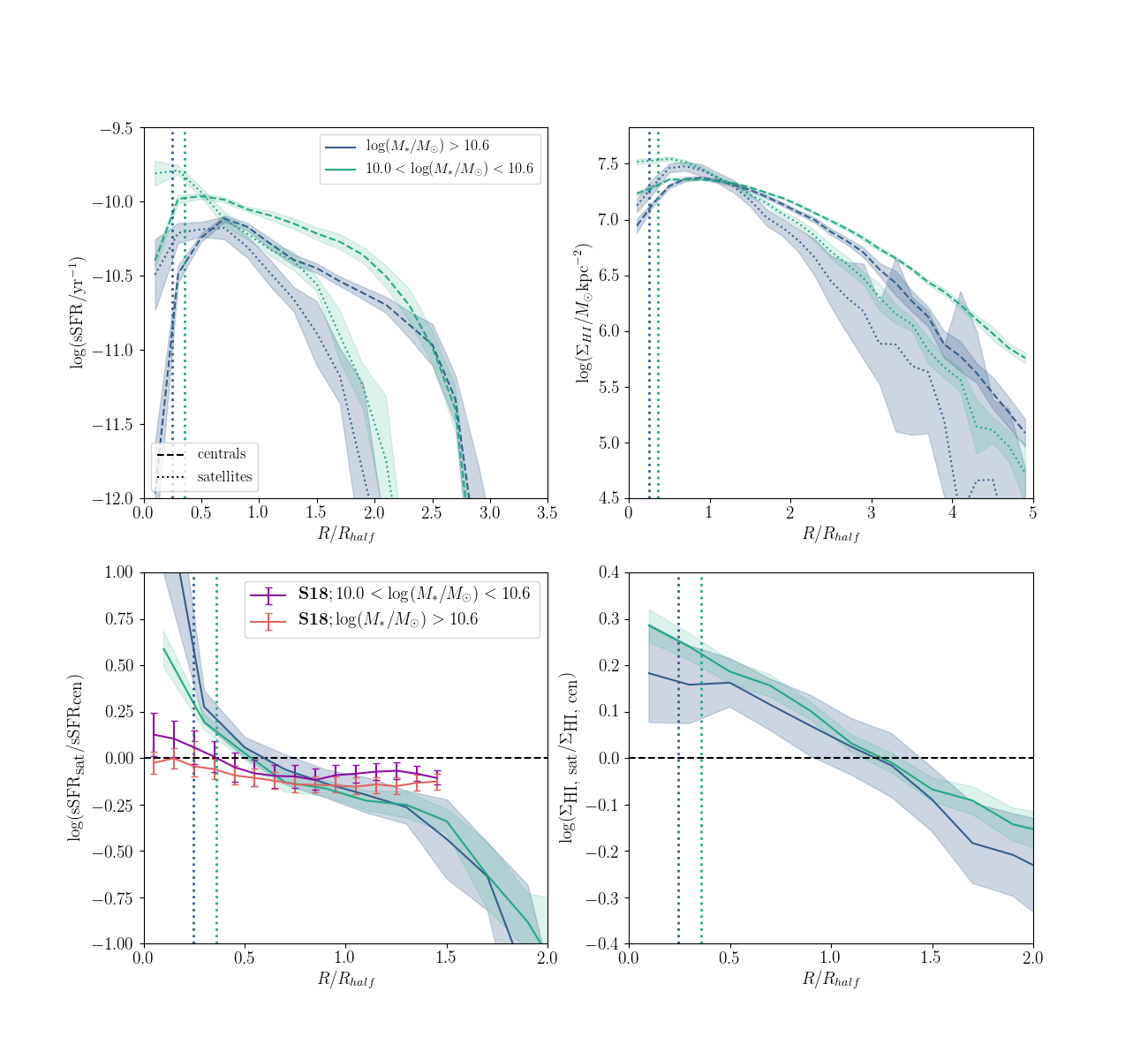}
	\vskip-0.4in
    \caption{{\it Top:} Radial profiles of $\Sigma_{\textrm{sSFR}}$ (left) and \(\Sigma_{\textrm{HI}}\) (right) for SF galaxies, split into centrals (dashed lines) and satellites (dotted lines), shown for two mass bins as green and blue lines. {\it Bottom:} Ratio of sSFR (left) and \(\Sigma_{\textrm{HI}}\) (right) radial profiles for satellite and central SF galaxies. Observations from \citet{spindler_2018} are shown as orange and purple lines. The horizontal black dashed line shows where the profiles for satellites and centrals are equal. For all panels, the displayed radial profiles are the ratio of the Tukey biweights of the satellite and central samples. The light shaded regions show the Tukey biweight scale estimator divided by \(\sqrt{N}\), combined with cosmic variance uncertainties from jackknife resampling over the 8 simulation sub-octants. \mycolor{The vertical dotted lines show the spatial resolution in units of the median half-light radius for each mass bin.}  Qualitatively, \simba reproduces the trends seen in the observations, suggesting that it accounts for both internal and external quenching processes in satellites. Quantitatively,  centrals in \simba have a significantly larger sSFR suppression than satellites at the same mass compared to observations, suggesting that X-ray AGN feedback is much weaker in satellites than in centrals.}
    \label{fig:sf_sat_cents}
\end{figure*}

Figure \ref{fig:sf_sat_cents} shows radial profiles of $\Sigma_{\textrm{sSFR}}$ and $\Sigma_{\HI}$ for galaxies separated into satellites and centrals (top row), and the logarithmic difference of the radial profiles (bottom row). Central galaxies are identified as the most baryonically massive galaxies in their halos, and satellites are all others.  We bin galaxies into two bins following \citet{spindler_2018}, namely $10\leq \log(\mstar/\msolar) \leq 10.6$ (turquoise, 1591 centrals and 349 satellites) and $\log(\mstar/\msolar) > 10.6$ (blue, 349 centrals and 74 satellites). They also have a bin to lower masses but we eschew this owing to numerical resolution concerns.  The \citet{spindler_2018} data in those bins is shown in orange and purple, respectively.

Qualitatively, \simba reproduces the trends seen in the observations. Beyond the core, satellites in \simba have lower sSFR than centrals, which is consistent with the data. The magnitude of the difference in the outer regions is similar to what is observed. Meanwhile, for $R\la 0.5\rhalf$, the trend reverses. This is broadly seen in the observations as well, though the trend does not fully reverse for the low-mass bin. This qualitative agreement suggests that \simba accounts for both internal and external quenching processes in satellites to yield rough agreement with observations.

Quantitatively, there are some significant differences. The bottom left panel shows that in \simba, centrals have a significantly larger sSFR suppression than satellites at the same mass, whereas the effect is relatively weak in the observations. This would suggest that X-ray AGN feedback is much weaker in satellites as in centrals, whereas it should be closer in order to match the data.  It is not immediately evident why X-ray feedback is weaker in satellites. Part of the difference could owe to differences in the way centrals and satellites are identified in observations versus our simulations; another possible explanation is that satellites may have preferentially lower mass black holes, which may make the transition to X-ray feedback less likely for these galaxies. We will explore this further in future work. The MaNGA data only probes out to $1.5\rhalf$, but \simba\ can examine these trends to larger radii, and predicts that the satellites truncate their sSFR at $\sim 2\rhalf$ as opposed to centrals which generally extend out to $\ga 3\rhalf$. 

Examining $\Sigma_{HI}(R)$ in the right panel, we see some differences between centrals and satellites but they are generally much reduced relative to that seen for sSFR. Satellites clearly show a steeper profile in the outskirts than centrals, but interestingly at $R\la \rhalf$ they actually have higher \HI\ surface densities.  A mild central depression is seen in \HI\ for the more massive galaxies, and this drop is identical in the centrals and satellites.  Hence \HI\ profiles are not quite as dramatically sensitive to environment as sSFR profiles, but nonetheless show a clear impact of gas suppression processes.

These trends are consistent with various potential environmental processes acting on satellite galaxies relative to centrals. As for why the outskirts of the centrals and satellites differ, satellites are more adversely affected by environmental processes as their lower masses leave outskirts more vulnerable. These processes include ram pressure stripping (removal of gas due to heating in the intracluster medium \citep{gunn_gott_1972}, galaxy harassment (gas removal due to frequent high speed galaxy encounters \citep{moore_1996}, mergers (collisions between galaxies \citep{toomre_1972} and strangulation (galaxies are unable to replenish their gas supply \citep{larson_1980, peng_2015}. These processes leave different observational signatures. Strangulation should deplete gas uniformly across the galaxy \citep[`anemic galaxies', e.g.][]{van_den_Burgh_1991, elmegreen_2002, spindler_2018}, whereas stripping removes gas preferentially from the outskirts and could lead to enhanced star formation confined to the galaxy centre \citep{spindler_2018}. In our case we see that the inner regions of the satellites are enhanced in both SFR and \HI, while the outskirts are more depleted than the centrals, which is broadly consistent with a ram pressure stripping scenario~\citep{cunnama_2014,rafieferantsoa_2019}.

\subsection{Redshift dependence}\label{sec:zevol}

We have shown that both inside-out and outside-in quenching occurs in \simba in $z=0$ GV galaxies, and that the driving physical mechanism within \simba\ appears to be its X-ray AGN feedback implementation.  An interesting question is, when do these quenching mechanisms become apparent?  At higher redshifts, it becomes more difficult to select GV galaxies, owing to overall younger stellar populations and an increased prevalence of dusty galaxies.  However, it is still possible to examine massive galaxies, which should have a higher fraction of galaxies in the process of quenching than at lower masses. Here we examine the radial sSFR profiles of SF galaxies at $z=2$ in \simba, as a function of \mstar, and compare to selected observations. 

\begin{figure}
	\includegraphics[width=\columnwidth]{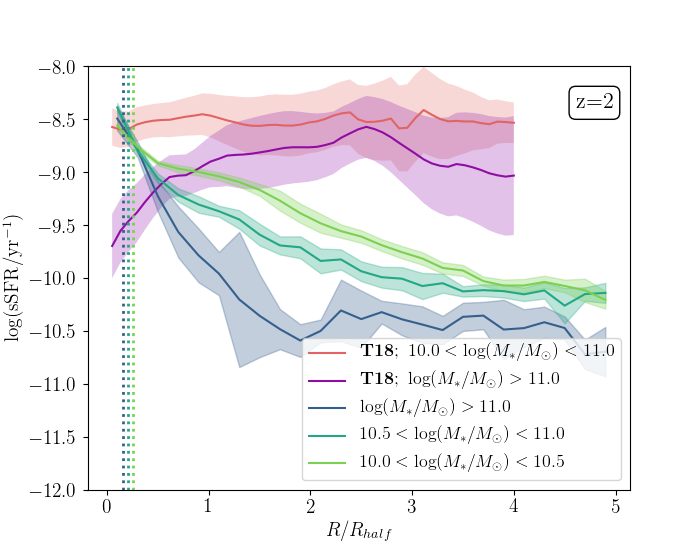}
	\vskip-0.1in
    \caption{Radial sSFR profiles for SF galaxies at \(z=2\). The displayed radial profiles are Tukey biweights of the individual galaxy profiles in each mass bin. The light shaded regions around each line show the Tukey biweight scale estimator divided by \(\sqrt{N}\), combined with cosmic variance uncertainties from jackknife resampling over the 8 simulation sub-octants. \mycolor{The vertical dotted lines show the spatial resolution in units of the median half-light radius for each mass bin.} Observations of sSFR profiles of SF galaxies from SINFONI \citep{tacchella_2018} at \(z=2\) are shown as the pink/purple lines. Active star formation at all radii across the galaxies suggests that neither the mechanisms for outside-in nor inside-out quenching have had a substantial impact. }
    \label{fig:redshift_ssfr}
\end{figure}

Figure \ref{fig:redshift_ssfr} shows the radial surface density profiles of sSFR for SF galaxies at $z = 2$. We select SF galaxies as having $\log \rm{(sSFR/yr}^{-1})$ > -9.5, to compare with SINFONI observations at $z=2$ \citep{tacchella_2018}, shown as orange/purple lines. We obtain 988 galaxies in the low-mass bin ($10^{10} \msolar <\mstar <10^{10.5} \msolar$), 318 in the intermediate-mass bin ($10^{10.5} \msolar <\mstar<10^{11} \msolar$), and 49 in the high-mass bin ($\mstar > 10^{11} \msolar$).

In general the sSFR profiles steadily increase towards the centre. At $z=2$, the profiles for low- and intermediate-mass bins are nearly identical, whereas the profile for the highest mass bin shows lower sSFR at all radii. The profile for the highest mass bin flattens at the centre, and the other masses show a slight decrease. 

Comparing with what we have already seen at low redshift, first we see that at $z=2$ there is active star formation at all radii across the galaxies, without a sharp decrease to zero at any point, whereas at low redshift the sSFR of SF galaxies drops to zero at $\sim 3 \rhalf$. Overall the level of star formation across the galaxies is higher, as is expected at high redshift. We find that the level of star formation is sensitive to the exact sSFR cut used to select the galaxies, but that the trends remain unchanged with different sSFR cuts, so in this case the shape of the profiles is more important than the exact level of star formation.

The trend of increasing star formation in the centre is similar to the {\it No-jet} and {\it No-X} models described in \S\ref{sec:feedback}, so it appears that the high redshift galaxies have not been affected by AGN feedback nearly to the same degree as the low redshift galaxies. This is not surprising since at high redshift \simba's AGN feedback has little impact in general. This is because the black holes at high redshift tend to have higher accretion rates and are thus emitting feedback radiatively (not in jet or X-ray form), which as shown in \citet{dave_2019} has little impact on galaxy properties. Similarly, at $z=2$ the environmental effects that cause the outside-in quenching (particularly in satellite galaxies, see section \ref{sec:cens_sats}) have not had a large impact, hence there is active star formation into the outskirts. It thus appears that at high redshift neither the mechanisms for outside-in nor inside-out quenching have had a substantial impact. In short, the AGN feedback modules responsible for quenching in \simba are not yet much in place at $z\sim 2$. 

In contrast, Figure \ref{fig:redshift_ssfr} shows that the observations indicate the more massive SF galaxies have clear central depressions in star formation even as early as $z=2$ \citep{tacchella_2018}. The observed sSFR profiles are flat for galaxies in the low mass bin, and the higher mass bin shows some suppression at the centre. It is clear that \simba does not fully reproduce the observed trends at high redshift, suggesting that some aspects of the AGN feedback implementation is unrealistic at early epochs. In particular, if X-ray feedback is responsible for creating central holes as seen at low redshift, then it may be that the X-ray feedback is insufficiently effective at higher $z$ in \simba. \mycolor{Inside-out quenching can also be associated with non-AGN related mechanisms, such as a wet compaction event due to minor mergers or tidal streams, which leads to a ring of star-forming gas around the centre \citep{tacchella_2016b}.}
Reproducing the central suppression in star formation at early epochs is thus another test that can be used to constrain AGN feedback models.

\subsection{\mycolor{Numerical convergence}}
\label{sec:numerical_convergence}

\begin{table*}
    \centering
    \caption{\mycolor{Median global galaxy properties and number of galaxies for the two mass bins for SF and GV galaxies in the high resolution simulation.}}
    \begin{tabular}{c|c c | c c }
        \hline
        Median & \multicolumn{2}{c}{Star forming} & \multicolumn{2}{c}{GV}  \\ 
        &  Low & Int & Low & Int \\
        \hline
        $N_{\textrm{gal}}$ & 61 & 31 & 35 & 11 \\
        $\textrm{log} (\mstar / \textrm{M}_{\odot})$ & 10.2 & 10.9 & 10.2 & 10.8 \\
        SFR (M$_{\odot} \textrm{yr}^{-1}$) & 1.53 & 5.48 & 0.19.& 0.78\\
        log (sSFR / yr$^{-1}$) & -10.06 & -10.25 & -10.89 & -10.81 \\
        $\textrm{log} (M_\HI / \textrm{M}_{\odot})$ & 10.2 & 10.4 & 9.6 & 9.9 \\
        $\textrm{log} (M_{\textrm{H}_2} / \textrm{M}_{\odot})$ & 9.4 & 9.9 & 8.6 & 9.2 \\
        $\rhalf$ (kpc) & 4.2 & 6.3 & 3.4 & 5.9 \\
        \hline
    \end{tabular}
    
    \label{tab:median_data_high_res}
\end{table*}

\begin{figure*}
	\vskip-0.1in
	\includegraphics[width=\textwidth]{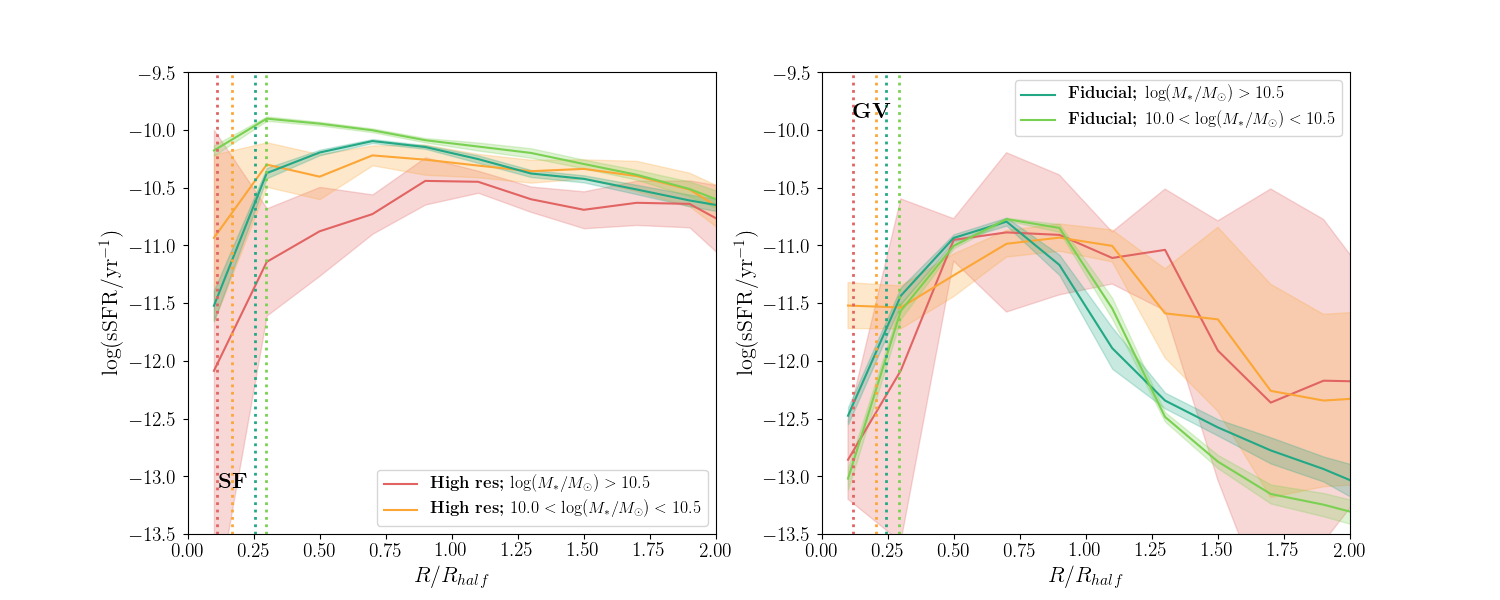}
	\vskip-0.15in
    \caption{\mycolor{Radial sSFR profiles for SF galaxies (left) and GV galaxies (right) in the high resolution Simba run (orange/purple lines) and the Simba run at fiducial resolution (green/blue lines). The vertical dotted lines show the spatial resolution in units of the median half-light radius for each mass bin, computed separately for SF and GV galaxies. The displayed radial profiles are Tukey biweights of the individual galaxy profiles in each mass bin. The light shaded regions around each line show the Tukey biweight scale estimator divided by \(\sqrt{N}\), combined with cosmic variance uncertainties from jackknife resampling over the 8 simulation sub-octants. Profiles of the high resolution galaxies are qualitatively similar to those from the main fiducial volume.}}
    \label{fig:high_res}
\end{figure*}

\mycolor{To test for numerical convergence, particularly in the most central radial bins, we compare to the higher resolution \simba\ volume of $25 \hmpc$ with $2\times512^3$ particles, which we denote \simba-hires. This simulation implements the same physical subgrid models as the main fiducial $100 \hmpc$ simulation, but with 8 times better mass resolution and twice the spatial resolution. }

\mycolor{We compute radial profiles of sSFR for SF and GV galaxies as before, using the same GV definition from \citet{belfiore_2018} as in section \ref{sec:belfiore_selection}. At $z=0$, we obtain 61 SF galaxies in the low-mass bin ($10^{10}\msolar <\mstar<10^{10.5}\msolar$) and 31 in the high-mass bin ($\mstar> 10^{10.5}\msolar$). In the same mass bins we find 35 and 11 GV galaxies respectively. With the small volume and high resolution, this simulation primarily probes galaxies of lower stellar mass, so the two highest mass bins are grouped together as there are few galaxies in these bins. Table \ref{tab:median_data_high_res} shows the median properties of SF and GV galaxies in this simulation.}

\mycolor{Figure \ref{fig:high_res} shows radial surface density profiles of sSFR for SF and GV galaxies in \simba-hires.  Owing to the low numbers, the profiles for the high resolution simulation are noisy, with a large range of sSFR per radial bin especially for high masses. In particular, there is a lot of variation in the centres of the SF galaxies and the outskirts of the GV galaxies.}

\mycolor{\simba-hires produces radial profiles that are qualitatively similar to those from the main \simba\ simulation. Importantly, the central suppression of star formation is still present in the high resolution case. The profiles for the high resolution GV galaxies show that the high mass galaxies show qualitatively the same level of star formation suppression as in the low resolution case. The SF galaxy profiles have the same shape as in the fiducial simulation, indicating that the quenching processes occur over the same radial scales as before. The broad agreement between the two simulations is an important result which reinforces the main trends described in this work and indicates that the qualitative trends are not sensitive to numerical resolution. }

\mycolor{Looking in more detail, there are some interesting discrepancies between the two simulations that point to the effects of numerical resolution. For the GV galaxies, inside-out quenching in low mass galaxies is not as strong in \simba-hires, indicating that these central radial bins were not well resolved in the low resolution case. The difference between the centres of the high and low resolution profiles is much more pronounced for the low galaxy mass bins, which intuitively makes sense as lower mass objects should be less well resolved to begin with. When analysing radial profiles elsewhere in this work, we should remember that the centres of the low mass GV galaxies are less well resolved. The low mass SF galaxies are more similar in terms of shape between the two simulations.}

\mycolor{In \S\ref{sec:sizemass} we noted that the sizes of passive galaxies are significantly smaller in \simba-hires, whereas the star forming galaxies were more numerically converged between the two simulations. The passive objects have few gas elements and are impacted by long-term numerical heating in the orbits of star particles that cannot dissipate their energy.  Hence star forming galaxies are expected to be better resolved generally. Following from this, profiles from star forming galaxies should be more robust than the green valley profiles as GV galaxies have lower gas fractions (see \S\ref{sec:radial}). This could explain why the inner regions of GV galaxies are less well converged than for SF galaxies, at least in terms of profile shape.}

\mycolor{For the GV galaxies, star formation extends further to the outskirts than for the low resolution profiles. The decrease in the outskirts is also more gradual than in the low resolution case -- outside-in quenching is somewhat weaker in the high resolution simulation. This could indicate that strong outside-in quenching is linked to low resolution. However, more likely it arises as a result of the small volume of the simulation.  In the $25\hmpc$ box, we do not produce the most massive galaxies with the strongest AGN feedback, potentially weakening the effect of environmental outside-in quenching.}

\mycolor{Looking at the SF galaxies, the high resolution case shows overall lower sSFR for both mass bins. This is a result of high galaxy stellar masses in the high resolution simulation owing to an abundance of lower mass star particles. The galaxy stellar mass function for the high resolution case does not match observations as well as the low resolution case. We have checked that profiles of SFR in the high resolution volume show that star formation activity is consistent with the main simulation.}

\mycolor{Overall, the results in this paper are qualitatively unchanged when examined at $8\times$ better mass resolution.  While this is still a limited dynamic range, it suggests that key results such as the suppression of gas and star formation in the central regions of GV galaxies is a robust outcome of the physics in \simba, rather than a numerical artifact.  Nonetheless, there is some non-trivial sensitivity to the results particularly for lower-mass galaxies that suggests caution when using such profiles to quantitatively constrain the input physics in \simba\ or simulations with similar resolution.}

\section{Summary}
\label{sec:summary}

We have examined the profiles of star-forming (SF) and GV (GV) galaxies in the \simba\ simulation, a state of the art cosmological hydrodynamic simulation that yields a population of quenched galaxies in good agreement with observations. \mycolor{We have examined the redshift evolution of half-light radii of SF and passive galaxies.} We separate SF and GV galaxies via a nonlinear cut in SFR--$\mstar$ space following \citet{belfiore_2018}, and focus on relatively well-resolved $\mstar>10^{10}M_\odot$ galaxies. We examine sSFR profiles, but also study profiles in SFR, gas surface density, star formation efficiency (or depletion time), and gas fraction.  We further examine differences in the profiles of central vs. satellite galaxies, and the evolution of sSFR profiles out to $z=2$.  We compare to $z=0$ observations of \citet{belfiore_2018} and \citet{spindler_2018}, and $z\sim 1-2$ data from \citet{tacchella_2018}.

Our main conclusions are as follows:

\begin{itemize}
    \item \simba\ reproduces $z=0$ star-forming galaxy sizes well, but yields quenched galaxies sizes that are too large at low $z$. \mycolor{The evolution of star forming sizes is also well reproduced, $\rhalf\propto(1+z)^{-0.78}$. However passive galaxies are of comparable size to star forming galaxies at $z=0$ and have too shallow a redshift evolution, $\propto(1+z)^{-1.06}$. This is consistent with the evolution of the numerical softening length, suggesting passive galaxy sizes are impacted by numerical resolution.}  For the majority of this paper, we only employ $z=0$ star-forming galaxy sizes.

    \item Examining $z=0$ galaxy profiles, we see that the surface density of star formation of all galaxies with $\mstar>10^{10}M_\odot$ in \simba\ shows a peak at $R\sim 0.5-1\rhalf$, where $\rhalf$ is the $r$-band half-light radius, an exponential dropoff to large radii, and a central $\Sigma_{SFR}$ depression in high-mass star forming galaxies and all GV galaxies. \mycolor{These trends at low radii are seen above the scale of the effective spatial resolution and broadly reproduces in a higher-resolution run, except for the case of the lowest mass galaxies.}
    
    \item The sSFR profile shows a qualitatively similar trend as the $\Sigma_{SFR}$ profile, with a sharper cutoff at $\sim 3\rhalf$.  Together, this shows that the central depression in the sSFR profile is a consequence of a lack of star formation in the core, not an excess of bulge stellar mass.
    
    \item GV galaxies show lower overall $\Sigma_{SFR}$ and sSFR at all radii, and have profiles with much larger central depressions and rapid truncation at $R\ga \rhalf$, than typical star-forming galaxies.
    
    \item The \HI\ surface density profiles for galaxies are virtually identical at $0.5\la R/\rhalf\la 3$ for all star-forming galaxies, but more massive systems show less $\Sigma_{HI}$ in the core and outskirts.  GV $\Sigma_{HI}$ profiles are similar to SF profiles at $R\la\rhalf$, but show a much more rapid decline beyond the this.
    
    \item The molecular gas profiles show considerably larger extent than the SFR or sSFR profiles, for both SF anf GV galaxies.  This is reflected in the H$_2$ fraction ($f_{H2}=\Sigma_{H2}/\Sigma_*$) profiles, which are fairly flat for $R\ga 0.5\rhalf$, for both galaxy types.  In the core region, $f_{H2}$ drops rapidly, showing that the central depression is caused by an evacuation of dense star-forming gas.  In the outer region, there is still substantial amounts of H$_2$, but it is evidently not forming stars.

    \item This is corroborated by examining the SF efficiency  (SFE=$\Sigma_{SFR}/\Sigma_{H2}$) profiles, which show a rapid decline in the outskirts but no central depression.  The change in GV profiles in the outskirts is thus entirely driven by a dropping SFE.  A simple scenario in which gas with $\Sigma_{H2}\la 1.1 M_\odot$pc$^{-2}$ doesn't form stars roughly reproduces the mean SFR truncation radius in SF vs. GV galaxies. 
    
    \item We compare our SF and GV sSFR profiles to observations of \citet{belfiore_2018}.  \simba\ yields a central depression in sSFR in qualitative agreement with data, unlike the EAGLE and Illustris simulations~\citep{starkenburg_2019}.  Quantitatively, however, massive ($\mstar>10^{10.5}M_\odot$) SF profiles show too large a central depression, and the shape of the central depression in the GV galaxies is not in perfect agreement with observations.
    
    \item  \mycolor{In contrast with observations}, \simba\ also yields a strong truncation in the sSFR profiles at $R\ga\rhalf$ for GV galaxies. This is present in all AGN feedback variants, so it is not associated in particular with quenching. In other words, \simba\ galaxies quench inside-out as observed, but some other physics may be incorrect which results in outside-in quenching; this may be related to the star formation prescription. 
    
    \item Using test simulations with various AGN feedback modules turned on and off, we demonstrate the it is specifically \simba's implementation of X-ray AGN feedback that is responsible for creating the central depression.  Turning this off results in steeply rising profiles for GV galaxies, as also seen in EAGLE and Illustris.  While \simba's X-ray feedback is quite heuristic (and was included mainly because of the physical motivation outlined in \citealt{choi_2012}), this demonstrates that some internal feedback process that generates outwards momentum deposition seems to be required in order to generate GV galaxy profiles as observed.
    
    \item Satellite galaxies show depressed sSFR relative to centrals at $R\ga\rhalf$, but an enhancement within (though still depressed overall).  This is qualitatively consistent with observations from \citet{spindler_2018}, though in \simba\ the core enhancement is larger than in the data.  \simba\ also shows a smaller radial extent of star formation in satellites vs. centrals, at radii beyond the range that is probed by the \citet{spindler_2018} data.
    
    \item The \HI\ surface density profile of satellites is likewise enhanced in the inner regions and depressed in the outer regions relative to centrals, but the effects are more modest than in the sSFR profiles.
    
    \item At $z=2$, \simba\ galaxies do not show central sSFR depressions in galaxies of any mass. This is understandable because X-ray feedback in \simba\ is tied to AGN jet feedback, which become widespread only at $z\la 1$.  However, this prediction is in contrast to observed sSFR profiles from \citet{tacchella_2018} at $z\sim 2$, which do show central depressions in the most massive SF galaxies.  Hence it appears that \simba's assumption of tying X-ray feedback to jet feedback may need to be revisited.
    
    \item \mycolor{The central star formation suppression is also produced in a \simba\ run with $8\times$ better mass resolution. GV galaxies in the high resolution volume show qualitatively similar levels of inside-out quenching, except in the inner radial bins of the lowest mass galaxies. Radial profiles of SF galaxies in the two simulations have the same shape, but the normalisation of the profile is low in the high resolution box, owing to more abundant low mass star particles. }
    
    \item \mycolor{Green valley galaxies in the high resolution simulation have more extended star formation. This is probably an outcome of the low volume of this simulation; the absence of the highest mass galaxies results in less heating of surrounding gas due to AGN feedback, and thus outside-in quenching due to the environment is weaker.}
\end{itemize}

Overall, our results demonstrate the valuable constraints provided by sSFR (and other) profiles of galaxies as they move from the star-forming to the quenched regime.  While it is encouraging that \simba's X-ray feedback reproduces the observed central sSFR depressions in GV galaxies, the various other discrepancies highlight that there are aspects of quenching in state of the art models that require further improvement.  Simulations such as \simba\ with sufficient resolution to examine the internal structure of galaxies, albeit coarsely, while still modeling a representative galaxy population, can now take advantage of these structural constraints to better understand the physical mechanisms by which galaxies quench.

\section*{Acknowledgements}

\mycolor{The authors thank the anonymous referee for their helpful comments and suggestions.} SA thanks Francesco Belfiore and Sandro Tacchella for providing their data. We thank Philip Hopkins for making \gizmo\ public, Robert Thompson for developing \caesar, and the YT team for development and support of YT. RD acknowledges support from the Wolfson Research Merit Award program of the U.K. Royal Society. DAA acknowledges support by the Flatiron Institute, which is supported by the Simons Foundation. DN was supported in part by National Science Foundation (NSF) Awards AST-1715206, AST-1909153, and AST-1908137, and HST Theory Award 15043.0001. Simba was run on the DiRAC@Durham facility managed by the Institute for Computational Cosmology on behalf of the Science and Technology Facilities Council (STFC) DiRAC HPC Facility. The equipment was funded by BEIS (Department for Business, Energy \& Industrial Strategy) capital funding via STFC capital grants ST/P002293/1, ST/R002371/1 and ST/S002502/1, Durham University and STFC operations grant ST/R000832/1. DiRAC is part of the National e-Infrastructure.

\bibliographystyle{mnras}
\bibliography{bib_main} 

\begin{thebibliography}{}
\makeatletter
\relax
\def\mn@urlcharsother{\let\do\@makeother \do\$\do\&\do\#\do\^\do\_\do\%\do\~}
\def\mn@doi{\begingroup\mn@urlcharsother \@ifnextchar [ {\mn@doi@}
  {\mn@doi@[]}}
\def\mn@doi@[#1]#2{\def\@tempa{#1}\ifx\@tempa\@empty \href
  {http://dx.doi.org/#2} {doi:#2}\else \href {http://dx.doi.org/#2} {#1}\fi
  \endgroup}
\def\mn@eprint#1#2{\mn@eprint@#1:#2::\@nil}
\def\mn@eprint@arXiv#1{\href {http://arxiv.org/abs/#1} {{\tt arXiv:#1}}}
\def\mn@eprint@dblp#1{\href {http://dblp.uni-trier.de/rec/bibtex/#1.xml}
  {dblp:#1}}
\def\mn@eprint@#1:#2:#3:#4\@nil{\def\@tempa {#1}\def\@tempb {#2}\def\@tempc
  {#3}\ifx \@tempc \@empty \let \@tempc \@tempb \let \@tempb \@tempa \fi \ifx
  \@tempb \@empty \def\@tempb {arXiv}\fi \@ifundefined
  {mn@eprint@\@tempb}{\@tempb:\@tempc}{\expandafter \expandafter \csname
  mn@eprint@\@tempb\endcsname \expandafter{\@tempc}}}

\bibitem[\protect\citeauthoryear{{Allen} et~al.,}{{Allen}
  et~al.}{2017}]{allen_2017}
{Allen} R.~J.,  et~al., 2017, \mn@doi [\apj] {10.3847/2041-8213/834/2/L11},
  \href {https://ui.adsabs.harvard.edu/abs/2017ApJ...834L..11A} {834, L11}

\bibitem[\protect\citeauthoryear{{Angl{\'e}s-Alc{\'a}zar}, {{\"O}zel}  \&
  {Dav{\'e}}}{{Angl{\'e}s-Alc{\'a}zar} et~al.}{2013}]{angles-alcazar_2013}
{Angl{\'e}s-Alc{\'a}zar} D.,  {{\"O}zel} F.,   {Dav{\'e}} R.,  2013, \mn@doi
  [\apj] {10.1088/0004-637X/770/1/5}, \href
  {https://ui.adsabs.harvard.edu/abs/2013ApJ...770....5A} {770, 5}

\bibitem[\protect\citeauthoryear{{Angl{\'e}s-Alc{\'a}zar}, {{\"O}zel},
  {Dav{\'e}}, {Katz}, {Kollmeier}  \& {Oppenheimer}}{{Angl{\'e}s-Alc{\'a}zar}
  et~al.}{2015}]{angles-alcazar_2015}
{Angl{\'e}s-Alc{\'a}zar} D.,  {{\"O}zel} F.,  {Dav{\'e}} R.,  {Katz} N.,
  {Kollmeier} J.~A.,   {Oppenheimer} B.~D.,  2015, \mn@doi [\apj]
  {10.1088/0004-637X/800/2/127}, \href
  {https://ui.adsabs.harvard.edu/abs/2015ApJ...800..127A} {800, 127}

\bibitem[\protect\citeauthoryear{{Angl{\'e}s-Alc{\'a}zar}, {Dav{\'e}},
  {Faucher-Gigu{\`e}re}, {{\"O}zel}  \& {Hopkins}}{{Angl{\'e}s-Alc{\'a}zar}
  et~al.}{2017a}]{angles-alcazar_2017a}
{Angl{\'e}s-Alc{\'a}zar} D.,  {Dav{\'e}} R.,  {Faucher-Gigu{\`e}re} C.-A.,
  {{\"O}zel} F.,   {Hopkins} P.~F.,  2017a, \mn@doi [\mnras]
  {10.1093/mnras/stw2565}, \href
  {https://ui.adsabs.harvard.edu/abs/2017MNRAS.464.2840A} {464, 2840}

\bibitem[\protect\citeauthoryear{{Angl{\'e}s-Alc{\'a}zar},
  {Faucher-Gigu{\`e}re}, {Kere{\v s}}, {Hopkins}, {Quataert}  \&
  {Murray}}{{Angl{\'e}s-Alc{\'a}zar} et~al.}{2017b}]{angles-alcazar_2017b}
{Angl{\'e}s-Alc{\'a}zar} D.,  {Faucher-Gigu{\`e}re} C.-A.,  {Kere{\v s}} D.,
  {Hopkins} P.~F.,  {Quataert} E.,   {Murray} N.,  2017b, \mn@doi [\mnras]
  {10.1093/mnras/stx1517}, \href
  {https://ui.adsabs.harvard.edu/abs/2017MNRAS.470.4698A} {470, 4698}

\bibitem[\protect\citeauthoryear{{Ascasibar}, {Guidi}, {Casado}, {Scannapieco}
  \& {D{\'\i}az}}{{Ascasibar} et~al.}{2016}]{ascasibar_2016}
{Ascasibar} Y.,  {Guidi} G.,  {Casado} J.,  {Scannapieco} C.,   {D{\'\i}az}
  A.~I.,  2016, arXiv e-prints, \href
  {https://ui.adsabs.harvard.edu/abs/2016arXiv160208474A} {p. arXiv:1602.08474}

\bibitem[\protect\citeauthoryear{{Baldry}, {Glazebrook}, {Brinkmann},
  {Ivezi{\'c}}, {Lupton}, {Nichol}  \& {Szalay}}{{Baldry}
  et~al.}{2004}]{baldry_2004}
{Baldry} I.~K.,  {Glazebrook} K.,  {Brinkmann} J.,  {Ivezi{\'c}} {\v{Z}}.,
  {Lupton} R.~H.,  {Nichol} R.~C.,   {Szalay} A.~S.,  2004, \mn@doi [\apj]
  {10.1086/380092}, \href
  {https://ui.adsabs.harvard.edu/abs/2004ApJ...600..681B} {600, 681}

\bibitem[\protect\citeauthoryear{{Balogh}, {Baldry}, {Nichol}, {Miller},
  {Bower}  \& {Glazebrook}}{{Balogh} et~al.}{2004}]{balogh_2004}
{Balogh} M.~L.,  {Baldry} I.~K.,  {Nichol} R.,  {Miller} C.,  {Bower} R.,
  {Glazebrook} K.,  2004, \mn@doi [\apjl] {10.1086/426079}, \href
  {https://ui.adsabs.harvard.edu/abs/2004ApJ...615L.101B} {615, L101}

\bibitem[\protect\citeauthoryear{{Belfiore} et~al.,}{{Belfiore}
  et~al.}{2018}]{belfiore_2018}
{Belfiore} F.,  et~al., 2018, \mn@doi [\mnras] {10.1093/mnras/sty768}, \href
  {https://ui.adsabs.harvard.edu/abs/2018MNRAS.477.3014B} {477, 3014}

\bibitem[\protect\citeauthoryear{{Bell} et~al.,}{{Bell}
  et~al.}{2004}]{bell_2004}
{Bell} E.~F.,  et~al., 2004, \mn@doi [\apj] {10.1086/420778}, \href
  {https://ui.adsabs.harvard.edu/abs/2004ApJ...608..752B} {608, 752}

\bibitem[\protect\citeauthoryear{{Bernardi}, {Meert}, {Vikram},
  {Huertas-Company}, {Mei}, {Shankar}  \& {Sheth}}{{Bernardi}
  et~al.}{2014}]{bernardi_2014}
{Bernardi} M.,  {Meert} A.,  {Vikram} V.,  {Huertas-Company} M.,  {Mei} S.,
  {Shankar} F.,   {Sheth} R.~K.,  2014, \mn@doi [\mnras]
  {10.1093/mnras/stu1106}, \href
  {https://ui.adsabs.harvard.edu/abs/2014MNRAS.443..874B} {443, 874}

\bibitem[\protect\citeauthoryear{{Bigiel} \& {Blitz}}{{Bigiel} \&
  {Blitz}}{2012}]{bigiel_2012}
{Bigiel} F.,  {Blitz} L.,  2012, \mn@doi [\apj] {10.1088/0004-637X/756/2/183},
  \href {https://ui.adsabs.harvard.edu/abs/2012ApJ...756..183B} {756, 183}

\bibitem[\protect\citeauthoryear{{Bigiel}, {Leroy}, {Walter}, {Brinks}, {de
  Blok}, {Madore}  \& {Thornley}}{{Bigiel} et~al.}{2008}]{bigiel_2008}
{Bigiel} F.,  {Leroy} A.,  {Walter} F.,  {Brinks} E.,  {de Blok} W.~J.~G.,
  {Madore} B.,   {Thornley} M.~D.,  2008, \mn@doi [\aj]
  {10.1088/0004-6256/136/6/2846}, \href
  {https://ui.adsabs.harvard.edu/abs/2008AJ....136.2846B} {136, 2846}

\bibitem[\protect\citeauthoryear{{Blanton}, {Kazin}, {Muna}, {Weaver}  \&
  {Price-Whelan}}{{Blanton} et~al.}{2011}]{blanton_2011}
{Blanton} M.~R.,  {Kazin} E.,  {Muna} D.,  {Weaver} B.~A.,   {Price-Whelan} A.,
   2011, \mn@doi [\aj] {10.1088/0004-6256/142/1/31}, \href
  {https://ui.adsabs.harvard.edu/abs/2011AJ....142...31B} {142, 31}

\bibitem[\protect\citeauthoryear{{Bondi}}{{Bondi}}{1952}]{bondi_1952}
{Bondi} H.,  1952, \mn@doi [\mnras] {10.1093/mnras/112.2.195}, \href
  {https://ui.adsabs.harvard.edu/abs/1952MNRAS.112..195B} {112, 195}

\bibitem[\protect\citeauthoryear{{Bower}, {Benson}, {Malbon}, {Helly}, {Frenk},
  {Baugh}, {Cole}  \& {Lacey}}{{Bower} et~al.}{2006}]{bower_2006}
{Bower} R.~G.,  {Benson} A.~J.,  {Malbon} R.,  {Helly} J.~C.,  {Frenk} C.~S.,
  {Baugh} C.~M.,  {Cole} S.,   {Lacey} C.~G.,  2006, \mn@doi [\mnras]
  {10.1111/j.1365-2966.2006.10519.x}, \href
  {https://ui.adsabs.harvard.edu/abs/2006MNRAS.370..645B} {370, 645}

\bibitem[\protect\citeauthoryear{{Bundy} et~al.,}{{Bundy}
  et~al.}{2010}]{bundy_2010}
{Bundy} K.,  et~al., 2010, \mn@doi [\apj] {10.1088/0004-637X/719/2/1969}, \href
  {https://ui.adsabs.harvard.edu/abs/2010ApJ...719.1969B} {719, 1969}

\bibitem[\protect\citeauthoryear{{Bundy} et~al.,}{{Bundy}
  et~al.}{2015}]{bundy_2015}
{Bundy} K.,  et~al., 2015, \mn@doi [\apj] {10.1088/0004-637X/798/1/7}, \href
  {https://ui.adsabs.harvard.edu/abs/2015ApJ...798....7B} {798, 7}

\bibitem[\protect\citeauthoryear{{Calzetti}, {Armus}, {Bohlin}, {Kinney},
  {Koornneef}  \& {Storchi-Bergmann}}{{Calzetti} et~al.}{2000}]{calzetti_2000}
{Calzetti} D.,  {Armus} L.,  {Bohlin} R.~C.,  {Kinney} A.~L.,  {Koornneef} J.,
   {Storchi-Bergmann} T.,  2000, \mn@doi [\apj] {10.1086/308692}, \href
  {https://ui.adsabs.harvard.edu/abs/2000ApJ...533..682C} {533, 682}

\bibitem[\protect\citeauthoryear{{Chabrier}}{{Chabrier}}{2003}]{chabrier_2003}
{Chabrier} G.,  2003, \mn@doi [\apjl] {10.1086/374879}, \href
  {https://ui.adsabs.harvard.edu/abs/2003ApJ...586L.133C} {586, L133}

\bibitem[\protect\citeauthoryear{{Choi}, {Ostriker}, {Naab}  \&
  {Johansson}}{{Choi} et~al.}{2012}]{choi_2012}
{Choi} E.,  {Ostriker} J.~P.,  {Naab} T.,   {Johansson} P.~H.,  2012, \mn@doi
  [\apj] {10.1088/0004-637X/754/2/125}, \href
  {https://ui.adsabs.harvard.edu/abs/2012ApJ...754..125C} {754, 125}

\bibitem[\protect\citeauthoryear{{Coenda}, {Mart{\'{\i}}nez}  \&
  {Muriel}}{{Coenda} et~al.}{2018}]{coenda_2018}
{Coenda} V.,  {Mart{\'{\i}}nez} H.~J.,   {Muriel} H.,  2018, \mn@doi [\mnras]
  {10.1093/mnras/stx2707}, \href
  {https://ui.adsabs.harvard.edu/abs/2018MNRAS.473.5617C} {473, 5617}

\bibitem[\protect\citeauthoryear{{Conroy} \& {Gunn}}{{Conroy} \&
  {Gunn}}{2010}]{conroy_2010}
{Conroy} C.,  {Gunn} J.~E.,  2010, \mn@doi [\apj]
  {10.1088/0004-637X/712/2/833}, \href
  {https://ui.adsabs.harvard.edu/abs/2010ApJ...712..833C} {712, 833}

\bibitem[\protect\citeauthoryear{{Conroy}, {Gunn}  \& {White}}{{Conroy}
  et~al.}{2009}]{conroy_2009}
{Conroy} C.,  {Gunn} J.~E.,   {White} M.,  2009, \mn@doi [\apj]
  {10.1088/0004-637X/699/1/486}, \href
  {https://ui.adsabs.harvard.edu/abs/2009ApJ...699..486C} {699, 486}

\bibitem[\protect\citeauthoryear{{Crain} et~al.,}{{Crain}
  et~al.}{2015}]{crain_2015}
{Crain} R.~A.,  et~al., 2015, \mn@doi [\mnras] {10.1093/mnras/stv725}, \href
  {https://ui.adsabs.harvard.edu/abs/2015MNRAS.450.1937C} {450, 1937}

\bibitem[\protect\citeauthoryear{{Croton} et~al.,}{{Croton}
  et~al.}{2006}]{croton_2006}
{Croton} D.~J.,  et~al., 2006, \mn@doi [\mnras]
  {10.1111/j.1365-2966.2005.09675.x}, \href
  {https://ui.adsabs.harvard.edu/abs/2006MNRAS.365...11C} {365, 11}

\bibitem[\protect\citeauthoryear{{Cunnama}, {Andrianomena}, {Cress},
  {Faltenbacher}, {Gibson}  \& {Theuns}}{{Cunnama} et~al.}{2014}]{cunnama_2014}
{Cunnama} D.,  {Andrianomena} S.,  {Cress} C.~M.,  {Faltenbacher} A.,  {Gibson}
  B.~K.,   {Theuns} T.,  2014, \mn@doi [\mnras] {10.1093/mnras/stt2380}, \href
  {https://ui.adsabs.harvard.edu/abs/2014MNRAS.438.2530C} {438, 2530}

\bibitem[\protect\citeauthoryear{{Dav{\'e}}, {Thompson}  \&
  {Hopkins}}{{Dav{\'e}} et~al.}{2016}]{dave_2016}
{Dav{\'e}} R.,  {Thompson} R.,   {Hopkins} P.~F.,  2016, \mn@doi [\mnras]
  {10.1093/mnras/stw1862}, \href
  {https://ui.adsabs.harvard.edu/abs/2016MNRAS.462.3265D} {462, 3265}

\bibitem[\protect\citeauthoryear{{Dav{\'e}}, {Angl{\'e}s-Alc{\'a}zar},
  {Narayanan}, {Li}, {Rafieferantsoa}  \& {Appleby}}{{Dav{\'e}}
  et~al.}{2019}]{dave_2019}
{Dav{\'e}} R.,  {Angl{\'e}s-Alc{\'a}zar} D.,  {Narayanan} D.,  {Li} Q.,
  {Rafieferantsoa} M.~H.,   {Appleby} S.,  2019, \mn@doi [\mnras]
  {10.1093/mnras/stz937}, \href
  {https://ui.adsabs.harvard.edu/abs/2019MNRAS.486.2827D} {486, 2827}

\bibitem[\protect\citeauthoryear{{Davies}, {Crain}, {Oppenheimer}  \&
  {Schaye}}{{Davies} et~al.}{2020}]{DaviesCrain2020}
{Davies} J.~J.,  {Crain} R.~A.,  {Oppenheimer} B.~D.,   {Schaye} J.,  2020,
  \mn@doi [\mnras] {10.1093/mnras/stz3201}, \href
  {https://ui.adsabs.harvard.edu/abs/2020MNRAS.491.4462D} {491, 4462}

\bibitem[\protect\citeauthoryear{{Davis}, {Greene}, {Ma}, {Pand ya},
  {Blakeslee}, {McConnell}  \& {Thomas}}{{Davis}
  et~al.}{2016}]{DavisGreeneMa2016}
{Davis} T.~A.,  {Greene} J.,  {Ma} C.-P.,  {Pand ya} V.,  {Blakeslee} J.~P.,
  {McConnell} N.,   {Thomas} J.,  2016, \mn@doi [\mnras]
  {10.1093/mnras/stv2313}, \href
  {https://ui.adsabs.harvard.edu/abs/2016MNRAS.455..214D} {455, 214}

\bibitem[\protect\citeauthoryear{{Dubois} et~al.,}{{Dubois}
  et~al.}{2014}]{dubois_2014}
{Dubois} Y.,  et~al., 2014, \mn@doi [\mnras] {10.1093/mnras/stu1227}, \href
  {http://adsabs.harvard.edu/abs/2014MNRAS.444.1453D} {444, 1453}

\bibitem[\protect\citeauthoryear{{Dubois}, {Peirani}, {Pichon}, {Devriendt},
  {Gavazzi}, {Welker}  \& {Volonteri}}{{Dubois} et~al.}{2016}]{dubois_2016}
{Dubois} Y.,  {Peirani} S.,  {Pichon} C.,  {Devriendt} J.,  {Gavazzi} R.,
  {Welker} C.,   {Volonteri} M.,  2016, \mn@doi [\mnras]
  {10.1093/mnras/stw2265}, \href
  {https://ui.adsabs.harvard.edu/abs/2016MNRAS.463.3948D} {463, 3948}

\bibitem[\protect\citeauthoryear{{Dwek}}{{Dwek}}{1998}]{dwek_1998}
{Dwek} E.,  1998, \mn@doi [\apj] {10.1086/305829}, \href
  {https://ui.adsabs.harvard.edu/abs/1998ApJ...501..643D} {501, 643}

\bibitem[\protect\citeauthoryear{{Ellison}, {S{\'a}nchez}, {Ibarra-Medel},
  {Antonio}, {Mendel}  \& {Barrera-Ballesteros}}{{Ellison}
  et~al.}{2018}]{ellison_2018}
{Ellison} S.~L.,  {S{\'a}nchez} S.~F.,  {Ibarra-Medel} H.,  {Antonio} B.,
  {Mendel} J.~T.,   {Barrera-Ballesteros} J.,  2018, \mn@doi [\mnras]
  {10.1093/mnras/stx2882}, \href
  {https://ui.adsabs.harvard.edu/abs/2018MNRAS.474.2039E} {474, 2039}

\bibitem[\protect\citeauthoryear{{Elmegreen}, {Elmegreen}, {Frogel},
  {Eskridge}, {Pogge}, {Gallagher}  \& {Iams}}{{Elmegreen}
  et~al.}{2002}]{elmegreen_2002}
{Elmegreen} D.~M.,  {Elmegreen} B.~G.,  {Frogel} J.~A.,  {Eskridge} P.~B.,
  {Pogge} R.~W.,  {Gallagher} A.,   {Iams} J.,  2002, \mn@doi [\aj]
  {10.1086/341613}, \href
  {https://ui.adsabs.harvard.edu/abs/2002AJ....124..777E} {124, 777}

\bibitem[\protect\citeauthoryear{{Faber} et~al.,}{{Faber}
  et~al.}{2007}]{faber_2007}
{Faber} S.~M.,  et~al., 2007, \mn@doi [\apj] {10.1086/519294}, \href
  {https://ui.adsabs.harvard.edu/abs/2007ApJ...665..265F} {665, 265}

\bibitem[\protect\citeauthoryear{{Fang}, {Faber}, {Salim}, {Graves}  \&
  {Rich}}{{Fang} et~al.}{2012}]{fang_2012}
{Fang} J.~J.,  {Faber} S.~M.,  {Salim} S.,  {Graves} G.~J.,   {Rich} R.~M.,
  2012, \mn@doi [\apj] {10.1088/0004-637X/761/1/23}, \href
  {https://ui.adsabs.harvard.edu/abs/2012ApJ...761...23F} {761, 23}

\bibitem[\protect\citeauthoryear{{Fitzpatrick} \& {Massa}}{{Fitzpatrick} \&
  {Massa}}{2007}]{fitzpatrick_massa_2007}
{Fitzpatrick} E.~L.,  {Massa} D.,  2007, \mn@doi [\apj] {10.1086/518158}, \href
  {https://ui.adsabs.harvard.edu/abs/2007ApJ...663..320F} {663, 320}

\bibitem[\protect\citeauthoryear{{Gabor} \& {Dav{\'e}}}{{Gabor} \&
  {Dav{\'e}}}{2012}]{gabor_2012}
{Gabor} J.~M.,  {Dav{\'e}} R.,  2012, \mn@doi [\mnras]
  {10.1111/j.1365-2966.2012.21640.x}, \href
  {https://ui.adsabs.harvard.edu/abs/2012MNRAS.427.1816G} {427, 1816}

\bibitem[\protect\citeauthoryear{{Gabor} \& {Dav{\'e}}}{{Gabor} \&
  {Dav{\'e}}}{2015}]{gabor_2015}
{Gabor} J.~M.,  {Dav{\'e}} R.,  2015, \mn@doi [\mnras] {10.1093/mnras/stu2399},
  \href {https://ui.adsabs.harvard.edu/abs/2015MNRAS.447..374G} {447, 374}

\bibitem[\protect\citeauthoryear{{Genel} et~al.,}{{Genel}
  et~al.}{2014}]{genel_2014}
{Genel} S.,  et~al., 2014, \mn@doi [\mnras] {10.1093/mnras/stu1654}, \href
  {https://ui.adsabs.harvard.edu/abs/2014MNRAS.445..175G} {445, 175}

\bibitem[\protect\citeauthoryear{{Genel} et~al.,}{{Genel}
  et~al.}{2018}]{genel_2018}
{Genel} S.,  et~al., 2018, \mn@doi [\mnras] {10.1093/mnras/stx3078}, \href
  {https://ui.adsabs.harvard.edu/abs/2018MNRAS.474.3976G} {474, 3976}

\bibitem[\protect\citeauthoryear{{Gonz{\'a}lez Delgado} et~al.,}{{Gonz{\'a}lez
  Delgado} et~al.}{2016}]{gonzalez_delgado_2016}
{Gonz{\'a}lez Delgado} R.~M.,  et~al., 2016, \mn@doi [\aap]
  {10.1051/0004-6361/201628174}, \href
  {https://ui.adsabs.harvard.edu/abs/2016A&A...590A..44G} {590, A44}

\bibitem[\protect\citeauthoryear{{Gunn} \& {Gott}}{{Gunn} \&
  {Gott}}{1972}]{gunn_gott_1972}
{Gunn} J.~E.,  {Gott} III J.~R.,  1972, \mn@doi [\apj] {10.1086/151605}, \href
  {https://ui.adsabs.harvard.edu/abs/1972ApJ...176....1G} {176, 1}

\bibitem[\protect\citeauthoryear{{Haardt} \& {Madau}}{{Haardt} \&
  {Madau}}{2012}]{haardt_2012}
{Haardt} F.,  {Madau} P.,  2012, \mn@doi [\apj] {10.1088/0004-637X/746/2/125},
  \href {https://ui.adsabs.harvard.edu/abs/2012ApJ...746..125H} {746, 125}

\bibitem[\protect\citeauthoryear{{Hahn} \& {Abel}}{{Hahn} \&
  {Abel}}{2011}]{hahn_2011}
{Hahn} O.,  {Abel} T.,  2011, \mn@doi [\mnras]
  {10.1111/j.1365-2966.2011.18820.x}, \href
  {https://ui.adsabs.harvard.edu/abs/2011MNRAS.415.2101H} {415, 2101}

\bibitem[\protect\citeauthoryear{{Heckman} \& {Best}}{{Heckman} \&
  {Best}}{2014}]{heckman_2014}
{Heckman} T.~M.,  {Best} P.~N.,  2014, \mn@doi [\araa]
  {10.1146/annurev-astro-081913-035722}, \href
  {https://ui.adsabs.harvard.edu/abs/2014ARA%26A..52..589H} {52, 589}

\bibitem[\protect\citeauthoryear{{Hopkins}}{{Hopkins}}{2015}]{hopkins_2015}
{Hopkins} P.~F.,  2015, \mn@doi [\mnras] {10.1093/mnras/stv195}, \href
  {https://ui.adsabs.harvard.edu/abs/2015MNRAS.450...53H} {450, 53}

\bibitem[\protect\citeauthoryear{{Hopkins} \& {Quataert}}{{Hopkins} \&
  {Quataert}}{2011}]{hopkins_2011a}
{Hopkins} P.~F.,  {Quataert} E.,  2011, \mn@doi [\mnras]
  {10.1111/j.1365-2966.2011.18542.x}, \href
  {https://ui.adsabs.harvard.edu/abs/2011MNRAS.415.1027H} {415, 1027}

\bibitem[\protect\citeauthoryear{{Hopkins}, {Cox}, {Kere{\v s}}  \&
  {Hernquist}}{{Hopkins} et~al.}{2008}]{hopkins_2008}
{Hopkins} P.~F.,  {Cox} T.~J.,  {Kere{\v s}} D.,   {Hernquist} L.,  2008,
  \mn@doi [\apjs] {10.1086/524363}, \href
  {https://ui.adsabs.harvard.edu/abs/2008ApJS..175..390H} {175, 390}

\bibitem[\protect\citeauthoryear{Hopkins, Kere\v{s}, O\~{n}orbe,
  Faucher-Gigu\`ere, Quataert, Murray  \& Bullock}{Hopkins
  et~al.}{2014}]{Hopkins_2014}
Hopkins P.~F.,  Kere\v{s} D.,  O\~{n}orbe J.,  Faucher-Gigu\`ere C.-A.,
  Quataert E.,  Murray N.,   Bullock J.~S.,  2014, \mn@doi [\mnras]
  {10.1093/mnras/stu1738}, 445, 581

\bibitem[\protect\citeauthoryear{{Kennicutt}}{{Kennicutt}}{1998}]{kennicutt_1998}
{Kennicutt} Jr. R.~C.,  1998, \mn@doi [\apj] {10.1086/305588}, \href
  {https://ui.adsabs.harvard.edu/abs/1998ApJ...498..541K} {498, 541}

\bibitem[\protect\citeauthoryear{{Kere{\v s}}, {Katz}, {Weinberg}  \&
  {Dav{\'e}}}{{Kere{\v s}} et~al.}{2005}]{keres_2005}
{Kere{\v s}} D.,  {Katz} N.,  {Weinberg} D.~H.,   {Dav{\'e}} R.,  2005, \mn@doi
  [\mnras] {10.1111/j.1365-2966.2005.09451.x}, \href
  {https://ui.adsabs.harvard.edu/abs/2005MNRAS.363....2K} {363, 2}

\bibitem[\protect\citeauthoryear{{Krumholz} \& {Gnedin}}{{Krumholz} \&
  {Gnedin}}{2011}]{krumholz_2011}
{Krumholz} M.~R.,  {Gnedin} N.~Y.,  2011, \mn@doi [\apj]
  {10.1088/0004-637X/729/1/36}, \href
  {https://ui.adsabs.harvard.edu/abs/2011ApJ...729...36K} {729, 36}

\bibitem[\protect\citeauthoryear{{Larson}, {Tinsley}  \& {Caldwell}}{{Larson}
  et~al.}{1980}]{larson_1980}
{Larson} R.~B.,  {Tinsley} B.~M.,   {Caldwell} C.~N.,  1980, \mn@doi [\apj]
  {10.1086/157917}, \href
  {https://ui.adsabs.harvard.edu/abs/1980ApJ...237..692L} {237, 692}

\bibitem[\protect\citeauthoryear{{Li}, {Narayanan}  \& {Dav{\'e}}}{{Li}
  et~al.}{2019}]{li_2019}
{Li} Q.,  {Narayanan} D.,   {Dav{\'e}} R.,  2019, \mn@doi [\mnras]
  {10.1093/mnras/stz2684}, \href
  {https://ui.adsabs.harvard.edu/abs/2019MNRAS.490.1425L} {490, 1425}

\bibitem[\protect\citeauthoryear{{Lin} et~al.,}{{Lin} et~al.}{2019}]{lin_2019}
{Lin} L.,  et~al., 2019, \mn@doi [\apj] {10.3847/1538-4357/aafa84}, \href
  {https://ui.adsabs.harvard.edu/abs/2019ApJ...872...50L} {872, 50}

\bibitem[\protect\citeauthoryear{{Liu} et~al.,}{{Liu} et~al.}{2018}]{liu_2018}
{Liu} F.~S.,  et~al., 2018, \mn@doi [\apj] {10.3847/1538-4357/aac20d}, \href
  {https://ui.adsabs.harvard.edu/abs/2018ApJ...860...60L} {860, 60}

\bibitem[\protect\citeauthoryear{{Ludlow}, {Schaye}, {Schaller}  \&
  {Richings}}{{Ludlow} et~al.}{2019}]{ludlow19}
{Ludlow} A.~D.,  {Schaye} J.,  {Schaller} M.,   {Richings} J.,  2019, \mn@doi
  [\mnras] {10.1093/mnrasl/slz110}, \href
  {https://ui.adsabs.harvard.edu/abs/2019MNRAS.488L.123L} {488, L123}

\bibitem[\protect\citeauthoryear{{Martin} et~al.,}{{Martin}
  et~al.}{2007}]{martin_2007}
{Martin} D.~C.,  et~al., 2007, \mn@doi [\apjs] {10.1086/516639}, \href
  {https://ui.adsabs.harvard.edu/abs/2007ApJS..173..342M} {173, 342}

\bibitem[\protect\citeauthoryear{{McKinnon}, {Torrey}  \&
  {Vogelsberger}}{{McKinnon} et~al.}{2016}]{mckinnon_2016}
{McKinnon} R.,  {Torrey} P.,   {Vogelsberger} M.,  2016, \mn@doi [\mnras]
  {10.1093/mnras/stw253}, \href
  {https://ui.adsabs.harvard.edu/abs/2016MNRAS.457.3775M} {457, 3775}

\bibitem[\protect\citeauthoryear{{McNamara} \& {Nulsen}}{{McNamara} \&
  {Nulsen}}{2007}]{mcnamara_2007}
{McNamara} B.~R.,  {Nulsen} P.~E.~J.,  2007, \mn@doi [\araa]
  {10.1146/annurev.astro.45.051806.110625}, \href
  {https://ui.adsabs.harvard.edu/abs/2007ARA%26A..45..117M} {45, 117}

\bibitem[\protect\citeauthoryear{{Mo}, {Mao}  \& {White}}{{Mo}
  et~al.}{1998}]{MoMaoWhite_98}
{Mo} H.~J.,  {Mao} S.,   {White} S.~D.~M.,  1998, \mn@doi [\mnras]
  {10.1046/j.1365-8711.1998.01227.x}, \href
  {https://ui.adsabs.harvard.edu/abs/1998MNRAS.295..319M} {295, 319}

\bibitem[\protect\citeauthoryear{{Moore}, {Katz}, {Lake}, {Dressler}  \&
  {Oemler}}{{Moore} et~al.}{1996}]{moore_1996}
{Moore} B.,  {Katz} N.,  {Lake} G.,  {Dressler} A.,   {Oemler} A.,  1996,
  \mn@doi [\nat] {10.1038/379613a0}, \href
  {https://ui.adsabs.harvard.edu/abs/1996Natur.379..613M} {379, 613}

\bibitem[\protect\citeauthoryear{{Nelson} et~al.,}{{Nelson}
  et~al.}{2016}]{nelson_2016}
{Nelson} E.~J.,  et~al., 2016, \mn@doi [\apj] {10.3847/0004-637X/828/1/27},
  \href {https://ui.adsabs.harvard.edu/abs/2016ApJ...828...27N} {828, 27}

\bibitem[\protect\citeauthoryear{{Nelson} et~al.,}{{Nelson}
  et~al.}{2019}]{nelson_2019}
{Nelson} D.,  et~al., 2019, \mn@doi [\mnras] {10.1093/mnras/stz2306}, \href
  {https://ui.adsabs.harvard.edu/abs/2019MNRAS.490.3234N} {490, 3234}

\bibitem[\protect\citeauthoryear{Orr et~al.,}{Orr et~al.}{2017}]{Orr_2017}
Orr M.~E.,  et~al., 2017, \mn@doi [The Astrophysical Journal]
  {10.3847/2041-8213/aa8f93}, 849, L2

\bibitem[\protect\citeauthoryear{{Oser}, {Naab}, {Ostriker}  \&
  {Johansson}}{{Oser} et~al.}{2012}]{oser_2012}
{Oser} L.,  {Naab} T.,  {Ostriker} J.~P.,   {Johansson} P.~H.,  2012, \mn@doi
  [\apj] {10.1088/0004-637X/744/1/63}, \href
  {https://ui.adsabs.harvard.edu/abs/2012ApJ...744...63O} {744, 63}

\bibitem[\protect\citeauthoryear{{Peng}, {Maiolino}  \& {Cochrane}}{{Peng}
  et~al.}{2015}]{peng_2015}
{Peng} Y.,  {Maiolino} R.,   {Cochrane} R.,  2015, \mn@doi [\nat]
  {10.1038/nature14439}, \href
  {https://ui.adsabs.harvard.edu/abs/2015Natur.521..192P} {521, 192}

\bibitem[\protect\citeauthoryear{{Perna}, {Lanzuisi}, {Brusa}, {Mignoli}  \&
  {Cresci}}{{Perna} et~al.}{2017}]{perna_2017}
{Perna} M.,  {Lanzuisi} G.,  {Brusa} M.,  {Mignoli} M.,   {Cresci} G.,  2017,
  \mn@doi [\aap] {10.1051/0004-6361/201630369}, \href
  {https://ui.adsabs.harvard.edu/abs/2017A%26A...603A..99P} {603, A99}

\bibitem[\protect\citeauthoryear{{Planck Collaboration} et~al.,}{{Planck
  Collaboration} et~al.}{2016}]{planck_collab_2016}
{Planck Collaboration} et~al., 2016, \mn@doi [\aap]
  {10.1051/0004-6361/201525830}, \href
  {https://ui.adsabs.harvard.edu/abs/2016A%26A...594A..13P} {594, A13}

\bibitem[\protect\citeauthoryear{{Quai}, {Pozzetti}, {Moresco}, {Citro},
  {Cimatti}, {Brinchmann}, {Gunawardhana}  \& {Paalvast}}{{Quai}
  et~al.}{2019}]{quai_2019}
{Quai} S.,  {Pozzetti} L.,  {Moresco} M.,  {Citro} A.,  {Cimatti} A.,
  {Brinchmann} J.,  {Gunawardhana} M. L.~P.,   {Paalvast} M.,  2019, \mn@doi
  [\mnras] {10.1093/mnras/stz2771}, \href
  {https://ui.adsabs.harvard.edu/abs/2019MNRAS.tmp.2361Q} {p.~2361}

\bibitem[\protect\citeauthoryear{{Rafieferantsoa}, {Dav{\'e}}  \&
  {Naab}}{{Rafieferantsoa} et~al.}{2019}]{rafieferantsoa_2019}
{Rafieferantsoa} M.,  {Dav{\'e}} R.,   {Naab} T.,  2019, \mn@doi [\mnras]
  {10.1093/mnras/stz1199}, \href
  {https://ui.adsabs.harvard.edu/abs/2019MNRAS.486.5184R} {486, 5184}

\bibitem[\protect\citeauthoryear{{Rahmati}, {Pawlik}, {Rai{\v c}evi{\'c}}  \&
  {Schaye}}{{Rahmati} et~al.}{2013}]{rahmati_2013}
{Rahmati} A.,  {Pawlik} A.~H.,  {Rai{\v c}evi{\'c}} M.,   {Schaye} J.,  2013,
  \mn@doi [\mnras] {10.1093/mnras/stt066}, \href
  {https://ui.adsabs.harvard.edu/abs/2013MNRAS.430.2427R} {430, 2427}

\bibitem[\protect\citeauthoryear{{Rodr{\'\i}guez Montero}, {Dav{\'e}}, {Wild},
  {Angl{\'e}s-Alc{\'a}zar}  \& {Narayanan}}{{Rodr{\'\i}guez Montero}
  et~al.}{2019}]{rodriguez-montero_2019}
{Rodr{\'\i}guez Montero} F.,  {Dav{\'e}} R.,  {Wild} V.,
  {Angl{\'e}s-Alc{\'a}zar} D.,   {Narayanan} D.,  2019, \mn@doi [\mnras]
  {10.1093/mnras/stz2580}, \href
  {https://ui.adsabs.harvard.edu/abs/2019MNRAS.490.2139R} {490, 2139}

\bibitem[\protect\citeauthoryear{{Salim} \& {Narayanan}}{{Salim} \&
  {Narayanan}}{2020}]{salim_narayanan_2020}
{Salim} S.,  {Narayanan} D.,  2020, arXiv e-prints, \href
  {https://ui.adsabs.harvard.edu/abs/2020arXiv200103181S} {p. arXiv:2001.03181}

\bibitem[\protect\citeauthoryear{{S{\'a}nchez} et~al.,}{{S{\'a}nchez}
  et~al.}{2018}]{sanchez_2018}
{S{\'a}nchez} S.~F.,  et~al., 2018, \rmxaa, \href
  {https://ui.adsabs.harvard.edu/abs/2018RMxAA..54..217S} {54, 217}

\bibitem[\protect\citeauthoryear{{Sanders} \& {Mirabel}}{{Sanders} \&
  {Mirabel}}{1996}]{sanders_1996}
{Sanders} D.~B.,  {Mirabel} I.~F.,  1996, \mn@doi [\araa]
  {10.1146/annurev.astro.34.1.749}, \href
  {https://ui.adsabs.harvard.edu/abs/1996ARA%26A..34..749S} {34, 749}

\bibitem[\protect\citeauthoryear{{Schawinski} et~al.,}{{Schawinski}
  et~al.}{2010}]{schawinski_2010}
{Schawinski} K.,  et~al., 2010, \mn@doi [\apj] {10.1088/0004-637X/711/1/284},
  \href {https://ui.adsabs.harvard.edu/abs/2010ApJ...711..284S} {711, 284}

\bibitem[\protect\citeauthoryear{{Schawinski} et~al.,}{{Schawinski}
  et~al.}{2014}]{schawinski_2014}
{Schawinski} K.,  et~al., 2014, \mn@doi [\mnras] {10.1093/mnras/stu327}, \href
  {https://ui.adsabs.harvard.edu/abs/2014MNRAS.440..889S} {440, 889}

\bibitem[\protect\citeauthoryear{{Schaye} et~al.,}{{Schaye}
  et~al.}{2015}]{schaye_2015}
{Schaye} J.,  et~al., 2015, \mn@doi [\mnras] {10.1093/mnras/stu2058}, \href
  {https://ui.adsabs.harvard.edu/abs/2015MNRAS.446..521S} {446, 521}

\bibitem[\protect\citeauthoryear{{Schmidt}}{{Schmidt}}{1959}]{schmidt_1959}
{Schmidt} M.,  1959, \mn@doi [\apj] {10.1086/146614}, \href
  {https://ui.adsabs.harvard.edu/abs/1959ApJ...129..243S} {129, 243}

\bibitem[\protect\citeauthoryear{{Shen}, {Mo}, {White}, {Blanton}, {Kauffmann},
  {Voges}, {Brinkmann}  \& {Csabai}}{{Shen} et~al.}{2003}]{shen_2003}
{Shen} S.,  {Mo} H.~J.,  {White} S.~D.~M.,  {Blanton} M.~R.,  {Kauffmann} G.,
  {Voges} W.,  {Brinkmann} J.,   {Csabai} I.,  2003, \mn@doi [\mnras]
  {10.1046/j.1365-8711.2003.06740.x}, \href
  {https://ui.adsabs.harvard.edu/abs/2003MNRAS.343..978S} {343, 978}

\bibitem[\protect\citeauthoryear{{Smith} et~al.,}{{Smith}
  et~al.}{2017}]{smith_2017}
{Smith} B.~D.,  et~al., 2017, \mn@doi [\mnras] {10.1093/mnras/stw3291}, \href
  {https://ui.adsabs.harvard.edu/abs/2017MNRAS.466.2217S} {466, 2217}

\bibitem[\protect\citeauthoryear{{Somerville} \& {Dav{\'e}}}{{Somerville} \&
  {Dav{\'e}}}{2015}]{somerville_2015}
{Somerville} R.~S.,  {Dav{\'e}} R.,  2015, \mn@doi [\araa]
  {10.1146/annurev-astro-082812-140951}, \href
  {https://ui.adsabs.harvard.edu/abs/2015ARA&A..53...51S} {53, 51}

\bibitem[\protect\citeauthoryear{{Somerville}, {Hopkins}, {Cox}, {Robertson}
  \& {Hernquist}}{{Somerville} et~al.}{2008}]{somerville_2008}
{Somerville} R.~S.,  {Hopkins} P.~F.,  {Cox} T.~J.,  {Robertson} B.~E.,
  {Hernquist} L.,  2008, \mn@doi [\mnras] {10.1111/j.1365-2966.2008.13805.x},
  \href {https://ui.adsabs.harvard.edu/abs/2008MNRAS.391..481S} {391, 481}

\bibitem[\protect\citeauthoryear{{Spilker}, {Bezanson}, {Weiner}, {Whitaker}
  \& {Williams}}{{Spilker} et~al.}{2019}]{spilker_2019}
{Spilker} J.~S.,  {Bezanson} R.,  {Weiner} B.~J.,  {Whitaker} K.~E.,
  {Williams} C.~C.,  2019, \mn@doi [\apj] {10.3847/1538-4357/ab3804}, \href
  {https://ui.adsabs.harvard.edu/abs/2019ApJ...883...81S} {883, 81}

\bibitem[\protect\citeauthoryear{{Spindler} et~al.,}{{Spindler}
  et~al.}{2018}]{spindler_2018}
{Spindler} A.,  et~al., 2018, \mn@doi [\mnras] {10.1093/mnras/sty247}, \href
  {https://ui.adsabs.harvard.edu/abs/2018MNRAS.476..580S} {476, 580}

\bibitem[\protect\citeauthoryear{{Springel}}{{Springel}}{2005}]{springel_2005}
{Springel} V.,  2005, \mn@doi [\mnras] {10.1111/j.1365-2966.2005.09655.x},
  \href {https://ui.adsabs.harvard.edu/abs/2005MNRAS.364.1105S} {364, 1105}

\bibitem[\protect\citeauthoryear{{Springel}, {Di Matteo}  \&
  {Hernquist}}{{Springel} et~al.}{2005}]{Springel_2005b}
{Springel} V.,  {Di Matteo} T.,   {Hernquist} L.,  2005, \mn@doi [\apjl]
  {10.1086/428772}, \href
  {https://ui.adsabs.harvard.edu/abs/2005ApJ...620L..79S} {620, L79}

\bibitem[\protect\citeauthoryear{{Starkenburg}, {Tonnesen}  \&
  {Kopenhafer}}{{Starkenburg} et~al.}{2019}]{starkenburg_2019}
{Starkenburg} T.~K.,  {Tonnesen} S.,   {Kopenhafer} C.,  2019, \mn@doi [\apjl]
  {10.3847/2041-8213/ab0f34}, \href
  {https://ui.adsabs.harvard.edu/abs/2019ApJ...874L..17S} {874, L17}

\bibitem[\protect\citeauthoryear{{Strateva} et~al.,}{{Strateva}
  et~al.}{2001}]{strateva_2001}
{Strateva} I.,  et~al., 2001, \mn@doi [\aj] {10.1086/323301}, \href
  {https://ui.adsabs.harvard.edu/abs/2001AJ....122.1861S} {122, 1861}

\bibitem[\protect\citeauthoryear{{Tacchella}, {Dekel}, {Carollo}, {Ceverino},
  {DeGraf}, {Lapiner}, {Mandelker}  \& {Primack}}{{Tacchella}
  et~al.}{2016}]{tacchella_2016b}
{Tacchella} S.,  {Dekel} A.,  {Carollo} C.~M.,  {Ceverino} D.,  {DeGraf} C.,
  {Lapiner} S.,  {Mandelker} N.,   {Primack} J.~R.,  2016, \mn@doi [\mnras]
  {10.1093/mnras/stw303}, \href
  {https://ui.adsabs.harvard.edu/abs/2016MNRAS.458..242T} {458, 242}

\bibitem[\protect\citeauthoryear{{Tacchella}, {Carollo}, {Dekel}, {Schreiber},
  {Renzini}  \& {zC-SINF Team}}{{Tacchella} et~al.}{2017}]{tacchella_2017}
{Tacchella} S.,  {Carollo} C.~M.,  {Dekel} A.,  {Schreiber} N.~F.,  {Renzini}
  A.,   {zC-SINF Team} 2017, in {Gil de Paz} A.,  {Knapen} J.~H.,   {Lee}
  J.~C.,  eds,  IAU Symposium Vol. 321, Formation and Evolution of Galaxy
  Outskirts. Cambridge University Press, pp 327--329,
  \mn@doi{10.1017/S1743921316011753}

\bibitem[\protect\citeauthoryear{{Tacchella} et~al.,}{{Tacchella}
  et~al.}{2018}]{tacchella_2018}
{Tacchella} S.,  et~al., 2018, \mn@doi [\apj] {10.3847/1538-4357/aabf8b}, \href
  {https://ui.adsabs.harvard.edu/abs/2018ApJ...859...56T} {859, 56}

\bibitem[\protect\citeauthoryear{{Thomas}, {Dav{\'e}}, {Angl{\'e}s-Alc{\'a}zar}
   \& {Jarvis}}{{Thomas} et~al.}{2019}]{thomas_2019}
{Thomas} N.,  {Dav{\'e}} R.,  {Angl{\'e}s-Alc{\'a}zar} D.,   {Jarvis} M.,
  2019, \mn@doi [\mnras] {10.1093/mnras/stz1703}, \href
  {https://ui.adsabs.harvard.edu/abs/2019MNRAS.487.5764T} {487, 5764}

\bibitem[\protect\citeauthoryear{{Toomre} \& {Toomre}}{{Toomre} \&
  {Toomre}}{1972}]{toomre_1972}
{Toomre} A.,  {Toomre} J.,  1972, \mn@doi [\apj] {10.1086/151823}, \href
  {https://ui.adsabs.harvard.edu/abs/1972ApJ...178..623T} {178, 623}

\bibitem[\protect\citeauthoryear{{Tsai} \& {Mathews}}{{Tsai} \&
  {Mathews}}{1995}]{tsai_mathews_1995}
{Tsai} J.~C.,  {Mathews} W.~G.,  1995, \mn@doi [\apj] {10.1086/175943}, \href
  {https://ui.adsabs.harvard.edu/abs/1995ApJ...448...84T} {448, 84}

\bibitem[\protect\citeauthoryear{{Vogelsberger} et~al.,}{{Vogelsberger}
  et~al.}{2014}]{vogelsberger_2014}
{Vogelsberger} M.,  et~al., 2014, \mn@doi [\mnras] {10.1093/mnras/stu1536},
  \href {https://ui.adsabs.harvard.edu/abs/2014MNRAS.444.1518V} {444, 1518}

\bibitem[\protect\citeauthoryear{{Watson}}{{Watson}}{2011}]{watson_2011}
{Watson} D.,  2011, \mn@doi [\aap] {10.1051/0004-6361/201117120}, \href
  {https://ui.adsabs.harvard.edu/abs/2011A&A...533A..16W} {533, A16}

\bibitem[\protect\citeauthoryear{{Weinberger} et~al.,}{{Weinberger}
  et~al.}{2018}]{weinberger_2018}
{Weinberger} R.,  et~al., 2018, \mn@doi [\mnras] {10.1093/mnras/sty1733}, \href
  {https://ui.adsabs.harvard.edu/abs/2018MNRAS.479.4056W} {479, 4056}

\bibitem[\protect\citeauthoryear{{Wild}, {Heckman}  \& {Charlot}}{{Wild}
  et~al.}{2010}]{wild_2010}
{Wild} V.,  {Heckman} T.,   {Charlot} S.,  2010, \mn@doi [\mnras]
  {10.1111/j.1365-2966.2010.16536.x}, \href
  {https://ui.adsabs.harvard.edu/abs/2010MNRAS.405..933W} {405, 933}

\bibitem[\protect\citeauthoryear{{Zabludoff}, {Zaritsky}, {Lin}, {Tucker},
  {Hashimoto}, {Shectman}, {Oemler}  \& {Kirshner}}{{Zabludoff}
  et~al.}{1996}]{zabludoff_1996}
{Zabludoff} A.~I.,  {Zaritsky} D.,  {Lin} H.,  {Tucker} D.,  {Hashimoto} Y.,
  {Shectman} S.~A.,  {Oemler} A.,   {Kirshner} R.~P.,  1996, \mn@doi [\apj]
  {10.1086/177495}, \href
  {https://ui.adsabs.harvard.edu/abs/1996ApJ...466..104Z} {466, 104}

\bibitem[\protect\citeauthoryear{{Zhang} \& {Yang}}{{Zhang} \&
  {Yang}}{2019}]{zhang_2019}
{Zhang} Y.-C.,  {Yang} X.-H.,  2019, \mn@doi [Research in Astronomy and
  Astrophysics] {10.1088/1674-4527/19/1/6}, \href
  {https://ui.adsabs.harvard.edu/abs/2019RAA....19....6Z} {19, 006}

\bibitem[\protect\citeauthoryear{van~den Burgh}{van~den
  Burgh}{1991}]{van_den_Burgh_1991}
van~den Burgh S.,  1991, \mn@doi [Publications of the Astronomical Society of
  the Pacific] {10.1086/132832}, 103, 390

\bibitem[\protect\citeauthoryear{{van der Wel} et~al.,}{{van der Wel}
  et~al.}{2014}]{van-der-wel_2014}
{van der Wel} A.,  et~al., 2014, \mn@doi [\apj] {10.1088/0004-637X/788/1/28},
  \href {https://ui.adsabs.harvard.edu/abs/2014ApJ...788...28V} {788, 28}

\makeatother
\end{thebibliography}

\bsp	
\label{lastpage}
\end{document}